\documentclass[twocolumn]{elsart3p}
\usepackage{graphicx} 
\usepackage{dcolumn}  
\usepackage{longtable}
\usepackage{bm}       
\usepackage{subfigure}
\usepackage{subeqn}
\begin{document}
\begin{frontmatter}
\title{An {\em Ad-Hoc} Method for Obtaining $\chi^2$ Values from Unbinned
Maximum Likelihood Fits}
\author{M. Williams} and
\author{C. A. Meyer} 
\address{Carnegie Mellon University, Pittsburgh PA, 15213}
\date{\today}

%
%
\begin{abstract} 
A common goal in an experimental physics analysis is to extract information
from a reaction with multi-dimensional kinematics. 
The preferred method for such a task is typically the unbinned maximum 
likelihood method. 
In fits using this method, 
the likelihood is a goodness-of-fit quantity in that it 
effectively discriminates between available hypotheses; however, it does 
not provide any information as to how well the best hypothesis describes the 
data. 
In this paper, we present an {\em ad-hoc} procedure for obtaining 
$\chi^2/n.d.f.$ values from unbinned maximum likelihood fits. This method does
not require binning the data, making it very applicable to multi-dimensional
problems.
\end{abstract}
\end{frontmatter}
%
\section{\label{section:intro}Introduction}
In many physics analyses, one seeks to extract information from a reaction 
with multi-dimensional kinematics. Fits to binned data are often limited by
statistical uncertainties; thus, use of the unbinned maximum likelihood method 
is generally preferred. This method is well suited to obtaining estimators
for unknown parameters in a hypothesis' probability density function (p.d.f.). 
It is also an excellent way of discriminating between hypotheses; however,
it does not provide a means of determining how well the best hypothesis 
describes the data. 

In this paper, we describe an {\em ad-hoc} procedure for obtaining a $\chi^2/n.d.f.$ 
goodness-of-fit measurement of a hypothesis whose parameters were determined using 
an unbinned maximum likelihood fit without having to bin the data. This quantity is 
not meant to replace the likelihood. The likelihood should still be used to determine 
which hypothesis best describes the data and to obtain estimators for any unknown 
parameters, {\em i.e.} extract any physical observables. The goodness-of-fit quantity 
obtained using our method would simply be used to judge how well the best hypothesis 
describes the data.

In our approach, event-by-event standardized residuals are used to obtain the global 
$\chi^2/n.d.f.$. These residuals can also be used for diagnostic purposes to study 
the goodness-of-fit as a function of kinematics or detector components. Determining 
where in phase space a fit fails to describe the data is vital to understanding what 
physics has not been accounted for in the p.d.f. Identifying regions or components of 
the detector where a fit fails to describe the data could help diagnose unaccounted 
for inefficiencies in the acceptance calculation.

Other authors have also developed methods for obtaining goodness-of-fit quantities
from unbinned data. The methods closest to ours are based on the nearest neighbors
to a given event~\cite{cite:friedman,cite:schilling1983,cite:schilling1986}. However,
there are several other recent articles on other procedures as well
\cite{cite:aslan,cite:kinoshita,cite:raja}. While these and other authors have developed 
methods for obtaining goodness-of-fit quantities from unbinned maximum likelihood 
fits, the advantage of our approach lies in the event-by-event residuals.  It is also 
straight-forward to apply our method to most experimental physics analyses.

\section{\label{section:method}The Method}
Consider a data set composed of $n$ total events, each of which is described 
by $m$ coordinates, $\vec{\xi}$.
The coordinates can be masses, angles, energies, {\em etc}.
Two separate hypotheses have been proposed to describe the data.
Their probability density functions will be denoted by
\begin{equation}
  \label{eq:method-pdf}
  F_h(\vec{\xi},\vec{\alpha}_h) = \frac{f_h(\vec{\xi},\vec{\alpha}_h)}
  {\int f_h(\vec{\xi},\vec{\alpha}_h) d\vec{\xi}},
\end{equation}
where $f_h$ are the functional dependencies on the coordinates, $\vec{\xi}$,
and (possible) unknown parameters, $\vec{\alpha}_h$, of hypothesis $h$ 
$(h = 1,2)$.

Using these p.d.f.'s, unbinned maximum likelihood fits can easily be performed
to obtain estimators, $\hat{\alpha}_h$, for the unknown parameters, 
$\vec{\alpha}_h$, of each hypothesis. The likelihoods from these fits,
$\mathcal{L}_h$, can be used to determine which hypothesis provides the
better description of the data; however, the following simple question about 
this hypothesis can not
yet be answered: ``How well does it describe the data?''

The aim of this procedure is to obtain a $\chi^2/n.d.f.$ 
goodness-of-fit measurement of
a hypothesis whose parameters were determined using an unbinned maximum 
likelihood fit. 
We first need to define a metric for the space spanned by the
``relevant'' coordinates of $\vec{\xi}$, {\em i.e.} all coordinates on which
the p.d.f has functional dependence.
A reasonable choice is to use 
$\delta_{kl}/\mathcal{R}^2_k$, where $\mathcal{R}_k$ is the maximum possible 
difference between the coordinates, $\xi_k$, of any two events in the sample. 
Using this metric, the distance between any two events, $d_{ij}$, 
is given by
\begin{equation}
  \label{eq:dist}
  d^2_{ij} = \sum\limits_{k} \left[ \frac{\xi^i_k - \xi^j_k}
    {\mathcal{R}_k} 
    \right]^2,
\end{equation}
where the sum is over all relevant coordinates (discussed above).

For each event, we then compute the distance, $r_i$, 
to the $n_c$'th closest event using (\ref{eq:dist}). The value of $n_c$ is
discussed below. Thus, a hypersphere centered at the $i^{th}$ event with radius
$r_i$ contains $n_c$ data events (excluding the event itself).
In this way, we can calculate the density of data events at the point 
$\vec{\xi}_i$. 
By comparing these density calculations to those predicted by a given
hypothesis, we can determine how well the hypothesis describes the data without
resorting to binning.

For each event, the standardized residual, $z$, for any hypothesis is then 
calculated according to
\begin{equation}
  \label{eq:chi2-event}
  z^2_i = \frac{(n_{m_i} - n_{p_i})^2}
      {\sigma^2_{m_i} + \sigma_{p_i}^2},
\end{equation}
where $n_{m_i}(n_{p_i})$ is the number of measured (predicted) events contained
within a hypersphere of radius $r_i$ centered at $\vec{\xi}_i$ and 
$\sigma_{m_i}(\sigma_{p_i})$ is the uncertainty on $n_{m_i}(n_{p_i})$.

For the case where the data set contains no background events
(dealing with backgrounds is discussed in Section~\ref{section:example:bkgd}), 
${n_{m_i}=n_c}$ and ${\sigma_{m_i}=\sqrt{n_c}}$, provided 
$n_c$ is chosen to be large enough for the Gaussian approximation to hold.
The number of predicted events, $n_{p_i}$, is calculated for hypothesis $h$ as
\begin{equation}
  \label{eq:method-num-predicted}
  n_{p_i} = n \left[\frac{
      \int_{\odot}
    f_h(\vec{\xi},\hat{\alpha}_h) d\vec{\xi}
  }
  {\int f_h(\vec{\xi},\hat{\alpha}_h) d\vec{\xi}}\right],
\end{equation}
where the integral in the numerator is over the hypersphere of radius $r_i$ 
centered at $\vec{\xi}_i$ discussed above. 

In principle, there are cases where the integrals in 
(\ref{eq:method-num-predicted}) can be performed analytically; however, once
detector acceptance is included, for which an 
analytic expression typically is not known, they must be done using the
Monte Carlo technique. For these cases, $n_{p}$
can be calculated by generating $n_{mc}$ Monte Carlo events according to 
$\vec{\xi}$ phase space which leads to the following approximation of 
(\ref{eq:method-num-predicted}):
\begin{equation}
  \label{eq:method-num-predicted-mc}
  n_{p_i} \approx n\left[\frac{
    \sum\limits_{j}^{n_{mc}} f_h(\vec{\xi}_j,\hat{\alpha}_h) \Theta(r_i - r_j)
  }
  {\sum\limits_{k}^{n_{mc}} f_h(\vec{\xi}_k,\hat{\alpha}_h)}\right],
\end{equation}
where the Heaviside function, $\Theta(r_i - r_j)$, is used to restrict the 
sum in the numerator to Monte Carlo events within the hypersphere.
The statistical uncertainty in the Monte Carlo calculation of $n_{p}$ would
be $\sigma_{p} = n_{p}/\sqrt{n^{r}_{mc}}$, where 
$n^{r}_{mc}$ is the number of Monte Carlo events within the hypersphere.
We note here that if $n^{r}_{mc}$ is small, then the uncertainty is better
approximated by $1+\sqrt{n^{r}_{mc} + 0.75}$~\cite{cite:small-numbers}.

If the hypothesis does in fact provide a good description of the data, then
the values of $z^2$ obtained from (\ref{eq:chi2-event}) will follow a 
$\chi^2$ distribution with one degree of freedom. The overall goodness-of-fit
can be obtained as follows:
\begin{equation}
  \chi^2/n.d.f. = \frac{1}{n-n_{par}}\sum\limits_{i}^{n} z^2_i,
\end{equation}
where $n_{par}$ is the number of free parameters present in the hypothesis'
p.d.f. Hypotheses which describe the data well will yield values of
$\chi^2/n.d.f. \approx 1$, while those which do not describe the data well
will yield larger values.

\section{\label{section:example}Example Application}

As an example, we will consider the reaction ${\gamma p \rightarrow p \omega}$
in a single $(s,t)$ bin, {\em i.e.} a single center-of-mass energy and 
production angle bin (extending the example to avoid binning in
production angle, or $t$, is discussed below). The $\omega$ decays to
$\pi^+\pi^-\pi^0$ about 90\% of the time; thus, we will assume we have a 
detector which has reconstructed ${\gamma p \rightarrow p \pi^+\pi^-\pi^0}$
events.  
Below we will analyze a simulated $\omega$ photoproduction data set. 
The goal of our model analysis is to extract the $\omega$ polarization
observables known as the {\em spin density matrix elements}, denoted by
$\rho^0_{MM'}$ (discussed below).

In terms of the mass of the $\pi^+\pi^-\pi^0$ system, $m_{3\pi}$, the $\omega$
events were generated according to 3-body phase space weighted by a 
Voigtian (a convolution of a Breit-Wigner and a Gaussian)
to account for both the natural width of the $\omega$ and 
detector resolution. For this example, we chose to use $\sigma = 5$~MeV/c$^2$
for the detector resolution. 
The goal of our analysis is to extract the three measurable elements of the
spin density matrix (for the case where neither the beam nor target are 
polarized) traditionally chosen to be $\rho^0_{00}$, $\rho^0_{1-1}$ and 
$Re\rho^0_{10}$. These can be accessed by examining the 
distribution of the decay products ($\pi^+\pi^-\pi^0$) of the $\omega$ in its
rest frame.

For this example, we chose to work in the helicity system which defines the
$z$ axis as the direction of the $\omega$ in the overall center-of-mass 
frame, the $y$ axis as the normal to the production plane and the $x$ axis is
simply given by ${\hat{x} = \hat{y} \times \hat{z}}$.
The decay angles $\theta,\phi$ are the polar and azimuthal angles of the normal
to the decay plane in the $\omega$ rest frame.
The decay angular distribution of the $\omega$ in this frame is then 
given by~\cite{cite:schilling-1970}
\begin{eqnarray}
  \label{eq:schil}
  W(\theta,\phi) 
  = \frac{3}{4\pi} \left(\frac{1}{2}(1 - \rho^0_{00})
  + \frac{1}{2}(3\rho^0_{00} - 1)\cos^2{\theta}\right. 
  \nonumber\\
  - \left.\rho^0_{1-1}\sin^2{\theta}\cos{2\phi}
  - \sqrt{2}Re\rho^0_{10}\sin{2\theta}\cos{\phi}\right).
\end{eqnarray}
We chose to use the values 
${\rho^0_{00} = 0.65}$, ${\rho^0_{1-1} = 0.05}$ and ${Re\rho^0_{10} = 0.10}$
for our simulated data set.

\subsection{\label{section:example:simple}No Background, 100\% Acceptance}

\begin{figure}[h!]
  \centering
  \subfigure[]{
  \includegraphics[width=0.45\textwidth]{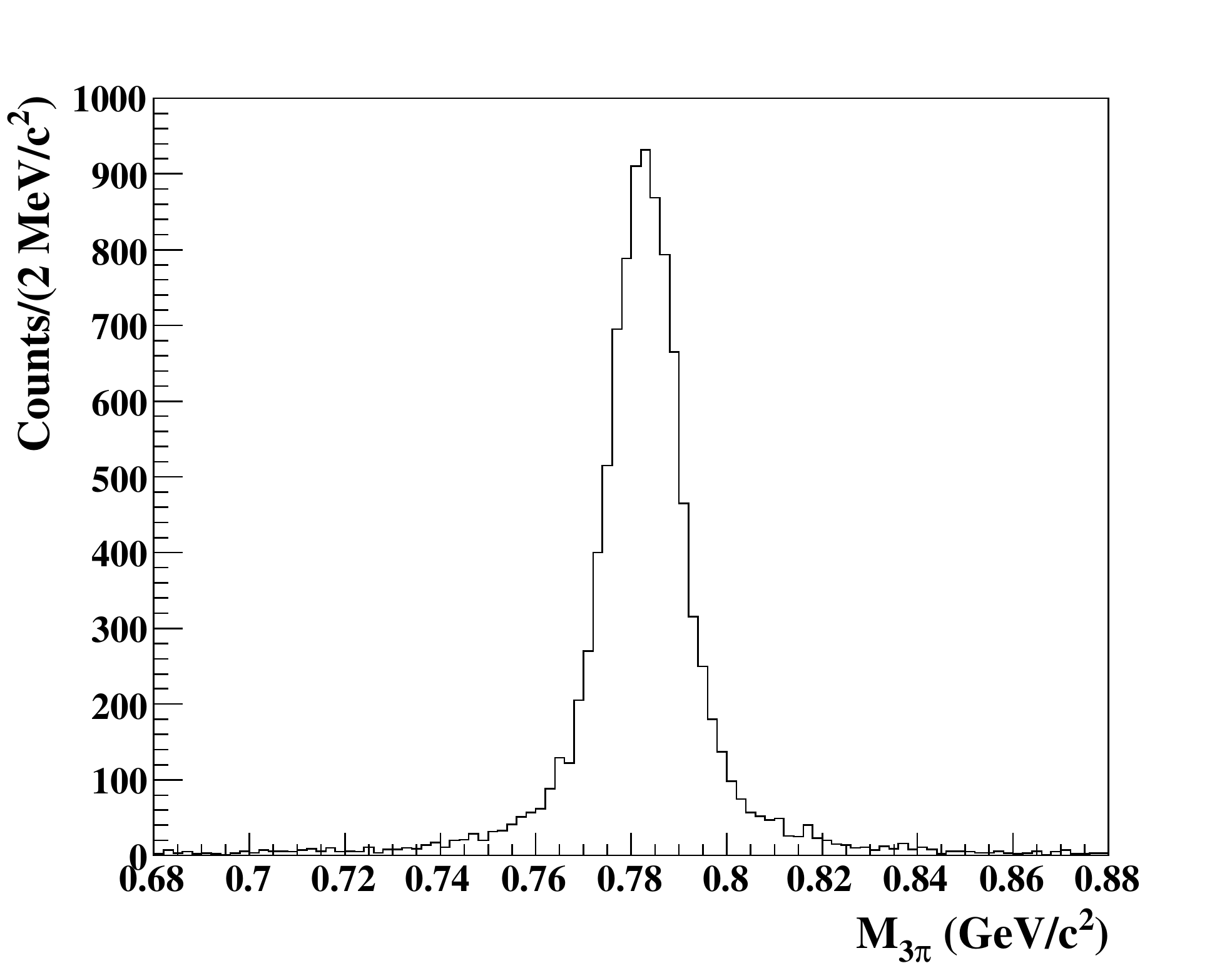}
  }
  \subfigure[]{
  \includegraphics[width=0.45\textwidth]{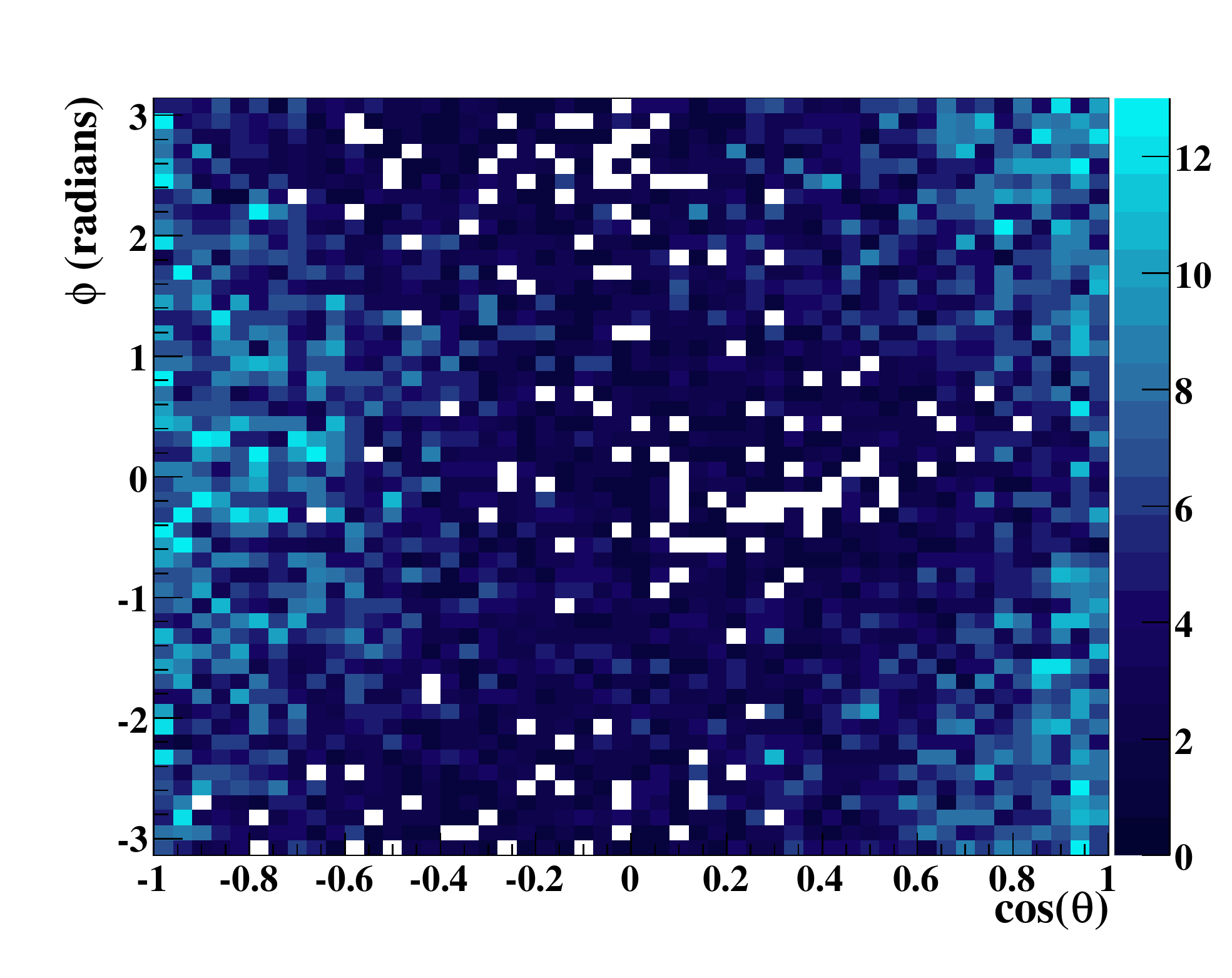}
  }
  \caption[]{\label{fig:gen-no-acc-no-bkgd}
    (Color Online) Events generated without background and with 100\% 
    acceptance. 
    (a) Mass of the $\pi^+\pi^-\pi^0$ system (GeV/c$^2$).
    (b) Decay angular distribution, $\phi$ (radians) vs $\cos{\theta}$.
  }
\end{figure}

We will begin by considering the simplest case where the signal is able to
be extracted without any background contamination and our detector efficiency
is 100\%. For this study, a simulated data set of 10,000 events was generated
following the criteria described above 
(see Figure~\ref{fig:gen-no-acc-no-bkgd}). 
To facilitate the calculation of
(\ref{eq:method-num-predicted-mc}), 100,000 Monte Carlo events were generated
using the same $m_{3\pi}$ distribution as the data; however, the decay 
distribution was generated according to flat phase space in $\cos{\theta}$ and 
$\phi$.

The likelihood function required to extract the spin density matrix elements 
is simply
\begin{equation}
  \mathcal{L} = \prod\limits_{i}^{n} W(\theta_i,\phi_i),
\end{equation}
where $W$ is the decay angular distribution defined in (\ref{eq:schil}). 
To obtain estimators for the $\rho^0_{MM'}$ elements, we minimize
\begin{equation}
  \label{eq:log-likelihood}
  -\ln{\mathcal{L}} = -\sum\limits_{i}^{n}\ln{W(\theta_i,\phi_i)},
\end{equation}
which yields the values
\begin{subequations}
  \label{eq:rho-no-acc-no-bkgd}
  \begin{equation}
  \rho^0_{00} = 0.649 \pm 0.009
  \end{equation}
  \begin{equation}  
  \rho^0_{1-1} = 0.043 \pm 0.005 
  \end{equation}
  \begin{equation}
    Re\rho^0_{10} = 0.103 \pm 0.004.
  \end{equation}
\end{subequations}
These values are in good agreement with the values used to generate the data.

At this point in our model analysis, we have extracted the physical observables
we are interested in. The spin and parity of the $\omega$ meson are well 
established; however, what if we were analyzing a less well-known meson? 
In that case, we may not be certain that $J^P = 1^-$, {\em i.e.} the use of 
(\ref{eq:schil}) might not be justified. Prior to publishing our results, it 
would be reassuring to know that the fit performed using (\ref{eq:schil})
does in fact provide a good description of our data.

\subsubsection{Applying the Procedure}

For this analysis, the relevant coordinates are the decay angles $\theta$ and
$\phi$, since these are the only kinematic quantities for which the p.d.f.
defined in  (\ref{eq:schil}) has dependence. The metric defined in 
(\ref{eq:dist}) is then given by
\begin{equation}
  \label{eq:example-metric}
  d^2_{ij} = \left(\frac{\cos{\theta_i} - \cos{\theta_j}}{2}\right)^2 
  + \left(\frac{\phi_i - \phi_j}{2\pi}\right)^2,
\end{equation}
for this analysis.

For each simulated event, we find the distance to the $n_c$'th nearest 
neighbor event, $r_i$. For now, we'll proceed using $n_c = 100$, determining
what to use for this value is discussed in detail in 
Section~\ref{section:example:nc}.
The phase space Monte Carlo events we generated are then used to obtain the
predicted number densities according to
\begin{equation}
  \label{eq:npi-no-bkgd-no-acc}
  n_{p_i} \approx n\left[\frac{
    \sum\limits_{j}^{n_{mc}} W(\theta_j,\phi_j) 
    \Theta(r_i - r_j)
  }
  {\sum\limits_{k}^{n_{mc}} W(\theta_k,\phi_k)}\right],
\end{equation}
using the $\rho^0_{MM'}$ values given in (\ref{eq:rho-no-acc-no-bkgd}). 
For this example, the 100\% acceptance makes explicitly computing the sum in
the denominator unnecessary; however, in the examples that follow it will be
required.

\begin{figure*}[p]
  \centering
  \subfigure[]{
  \includegraphics[width=0.3\textwidth]{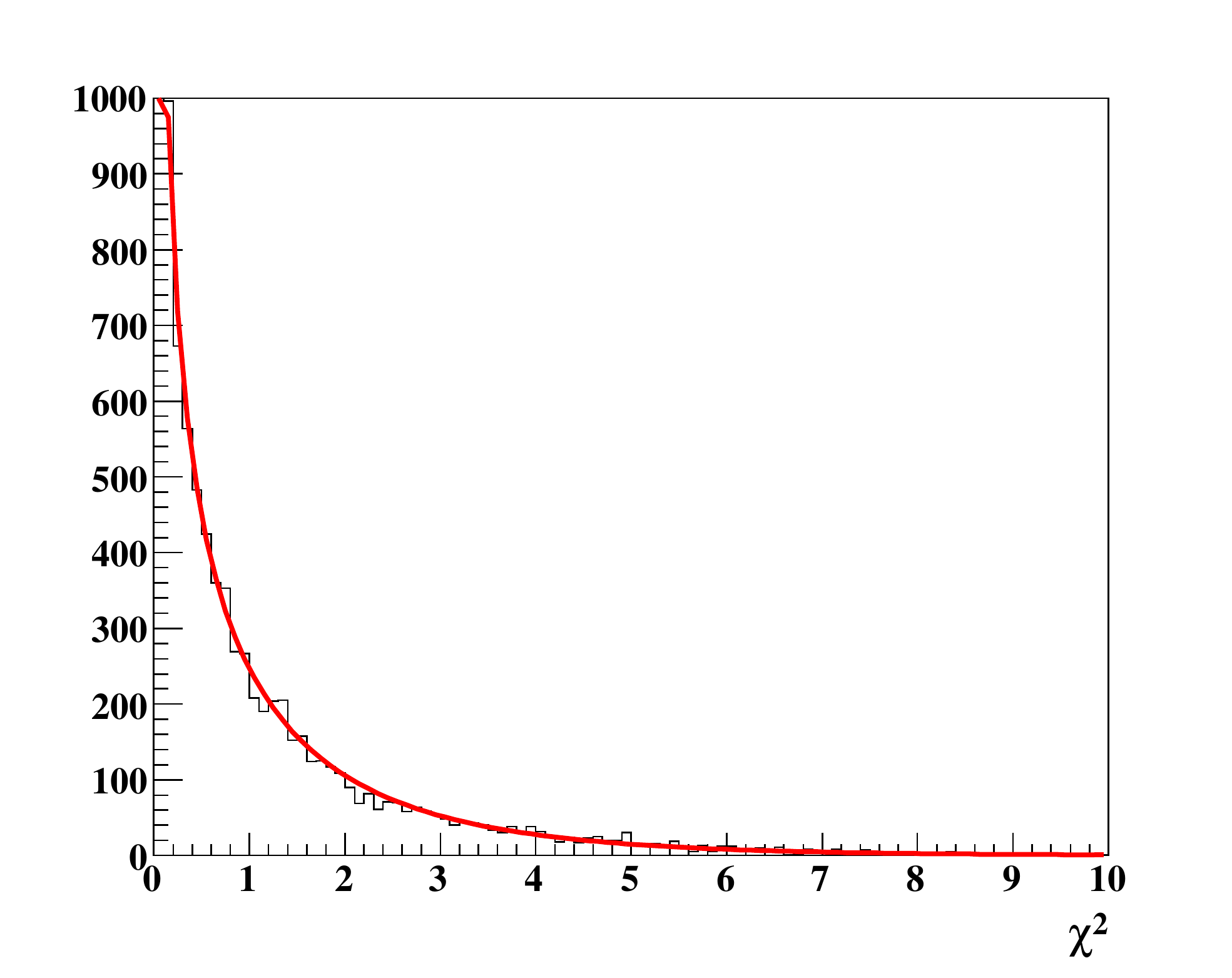}
  }
  \subfigure[]{
  \includegraphics[width=0.3\textwidth]{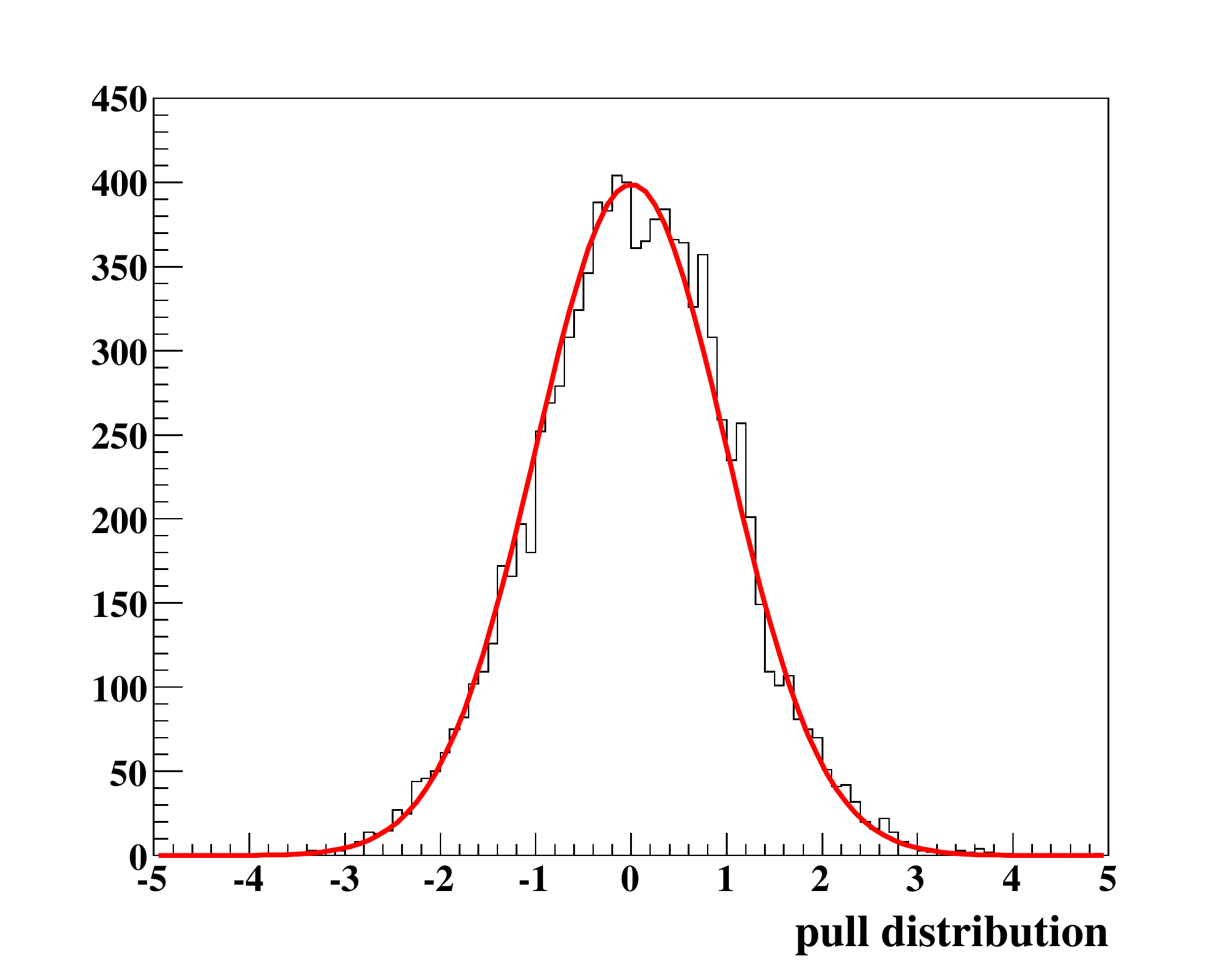}
  }
  \subfigure[]{
  \includegraphics[width=0.3\textwidth]{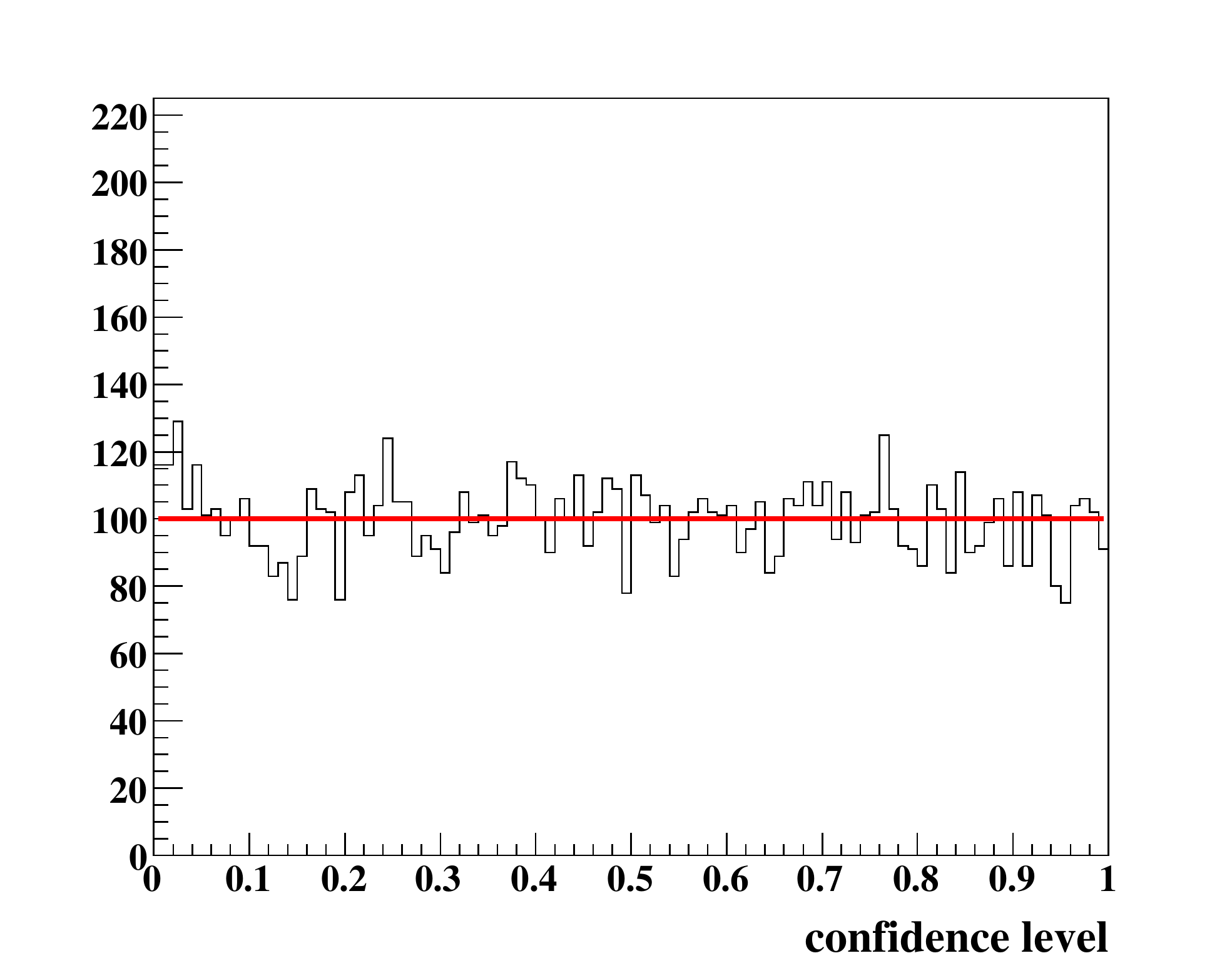}
  }
  \caption[]{\label{fig:chi2-no-acc-no-bkgd}
    (Color Online) Events generated without background and with 100\% 
    acceptance. The plots represent the
    (a) $\chi^2$ 
    (b) pull and 
    (c) confidence level distributions obtained from a fit to (\ref{eq:schil}).
    The red lines represent the theoretical distributions, which contain no
    free parameters.
    See text for details.
  }
\end{figure*}

The standardized residuals, $z$, for this hypothesis are obtained from 
(\ref{eq:chi2-event}) as
\begin{equation}
  z^2_i = \frac{(100 - n_{p_i})^2}{100 + n_{p_i}},
\end{equation}
where $n_{p_i}$ is obtained from (\ref{eq:npi-no-bkgd-no-acc}).
Figure~\ref{fig:chi2-no-acc-no-bkgd} shows the $\chi^2$, pull and confidence
level distributions obtained for our simulated data set. The values are in
excellent agreement with the theoretical distributions, {\em i.e.} they do
follow a $\chi^2$ distribution.
The total $\chi^2$ is obtained by summing over all events.  For this example,
the value is 10344; thus,
$\chi^2/n.d.f. = 10344/(10000-3) = 1.035$; 
a clear indicator that this hypothesis provides an excellent description of
the data.

\begin{figure*}[p]
  \centering
  \subfigure[]{
  \includegraphics[width=0.3\textwidth]{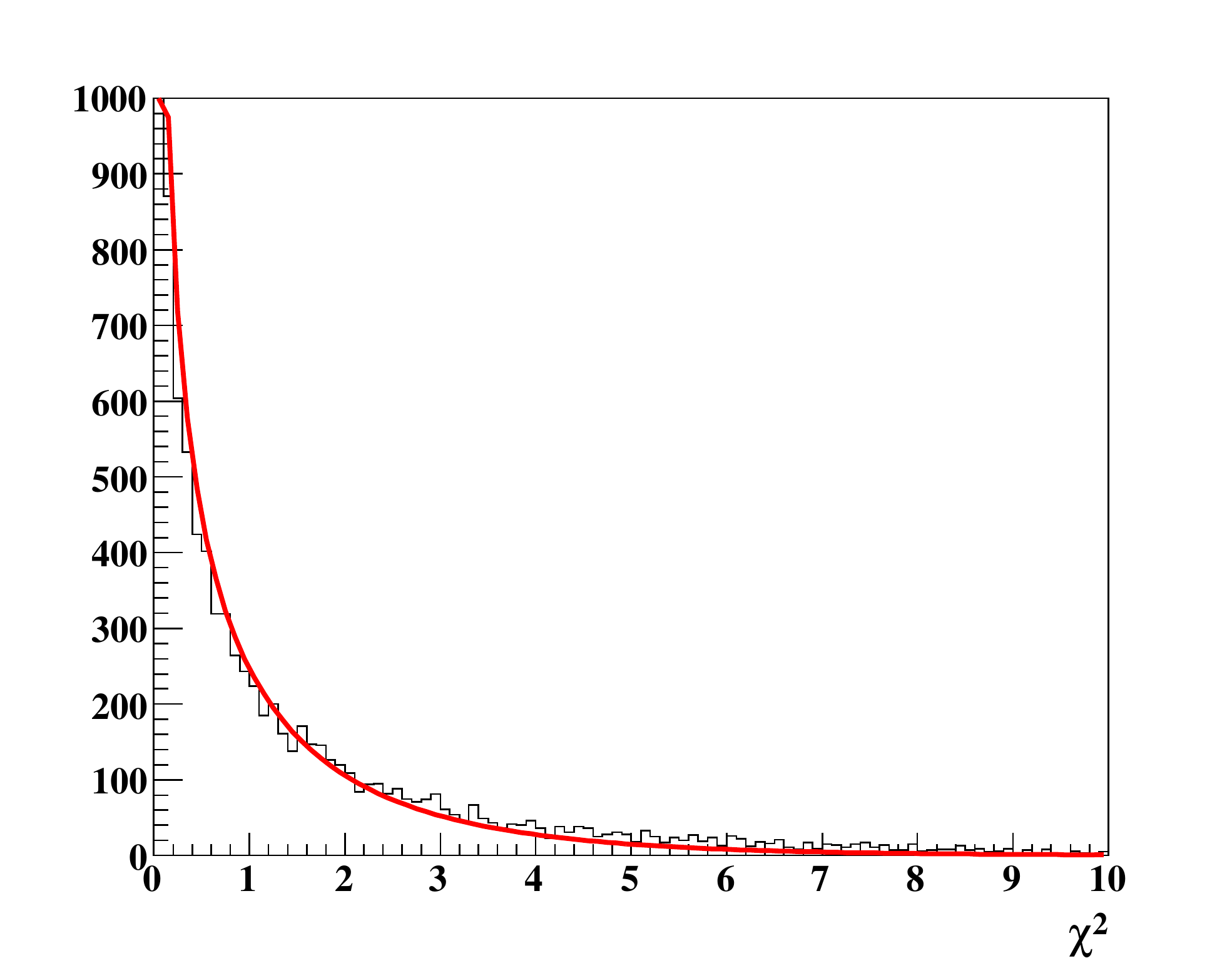}
  }
  \subfigure[]{
  \includegraphics[width=0.3\textwidth]{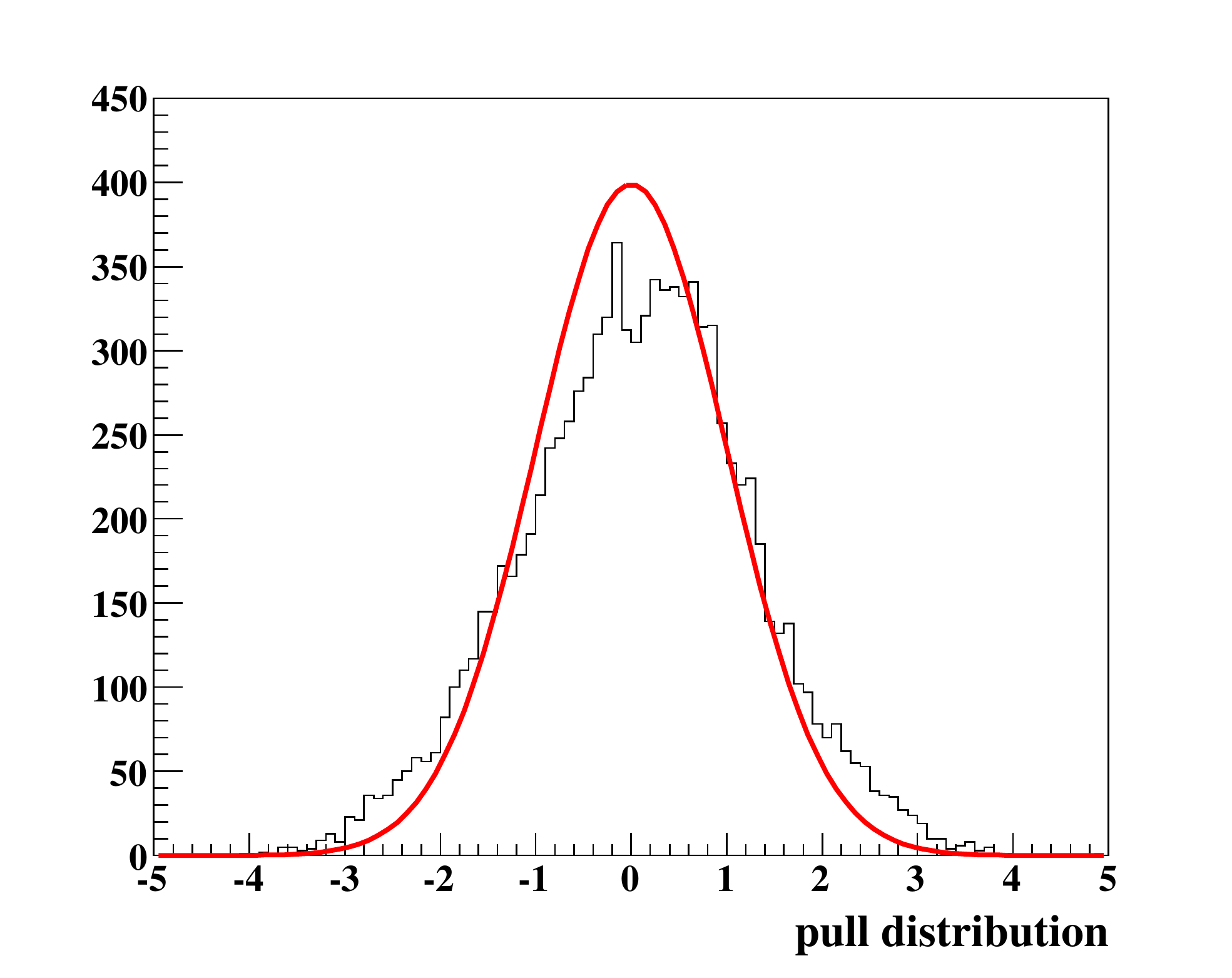}
  }
  \subfigure[]{
  \includegraphics[width=0.3\textwidth]{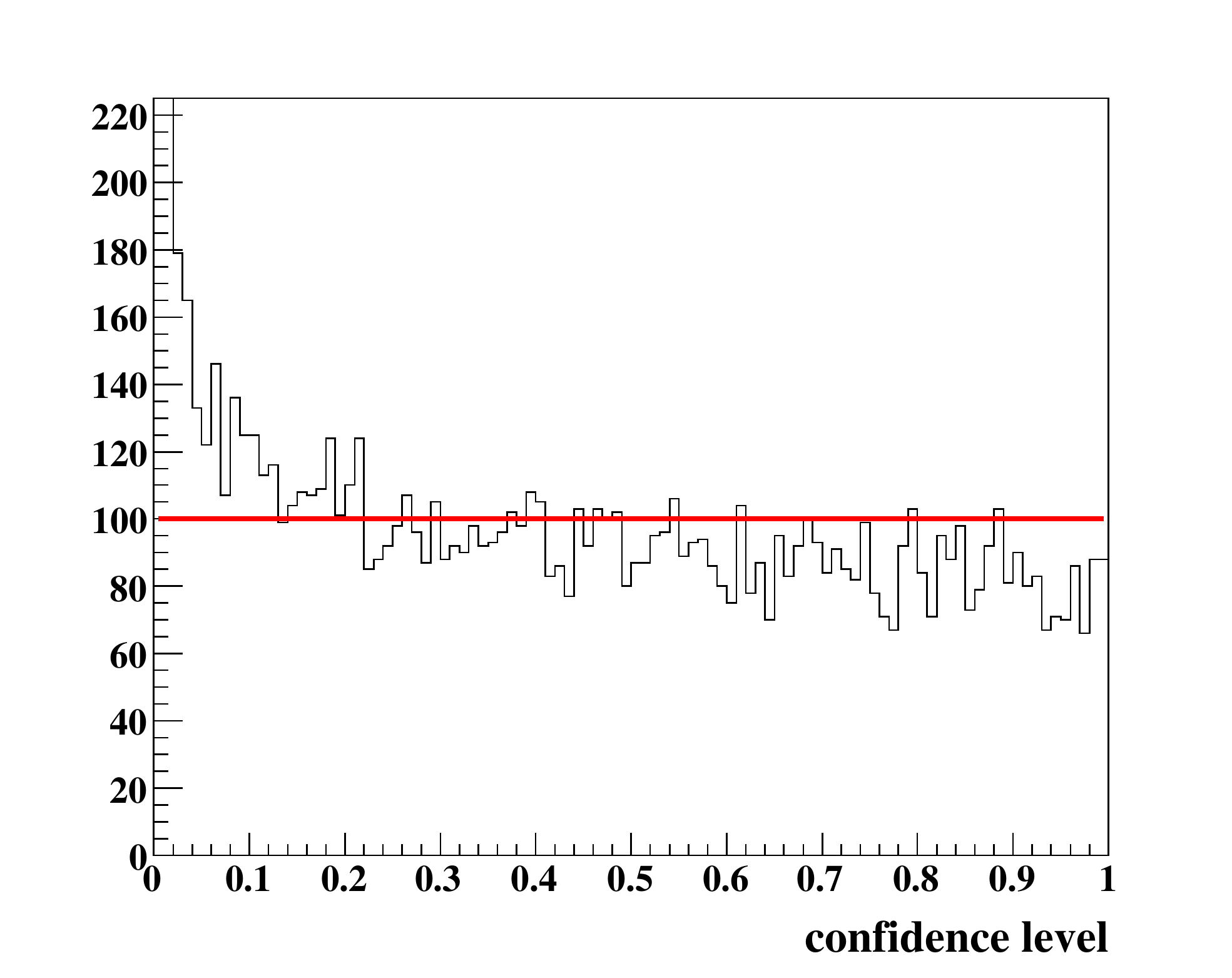}
  }
  \caption[]{\label{fig:chi2-no-1-1-no-acc-no-bkgd}
    (Color Online) Events generated without background and with 100\% 
    acceptance. The plots represent the
    (a) $\chi^2$ 
    (b) pull and 
    (c) confidence level distributions obtained from a fit to (\ref{eq:schil})
    in which the $\rho^0_{1-1}$ off-diagonal element was constrained to be 
    zero. 
    The red lines represent the theoretical distributions, which contain no
    free parameters.
    See text for details.
  }
\end{figure*}

It is also instructive to examine hypotheses which do not perfectly describe
the data. If the likelihood is maximized while requiring 
$\rho^0_{1-1} = 0$, then the spin density matrix elements extracted are 
${\rho^0_{00} = 0.650\pm0.009}$ and ${Re\rho^0_{10} = 0.110\pm0.004}$.
Figure~\ref{fig:chi2-no-1-1-no-acc-no-bkgd} shows the $\chi^2$, pull and 
confidence level distributions obtained for our simulated data set under this
hypothesis. 
Since the data were generated with ${\rho^0_{1-1} = 0.05}$, this fit provides a
fair description of the data;
the total $\chi^2/n.d.f. = 14527/(10000-2) = 1.453$. 

\begin{figure*}[p]
  \centering
  \subfigure[]{
  \includegraphics[width=0.3\textwidth]{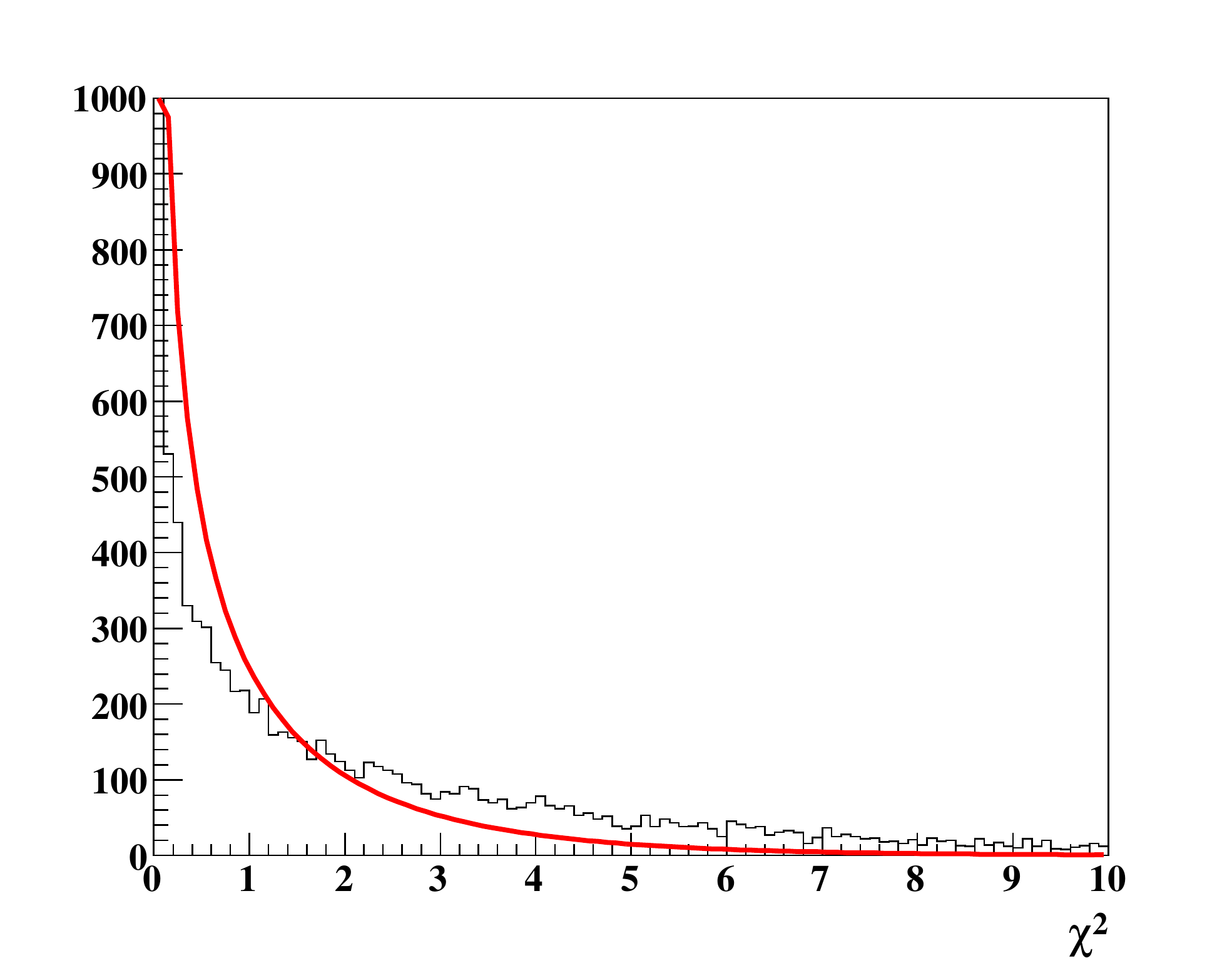}
  }
  \subfigure[]{
  \includegraphics[width=0.3\textwidth]{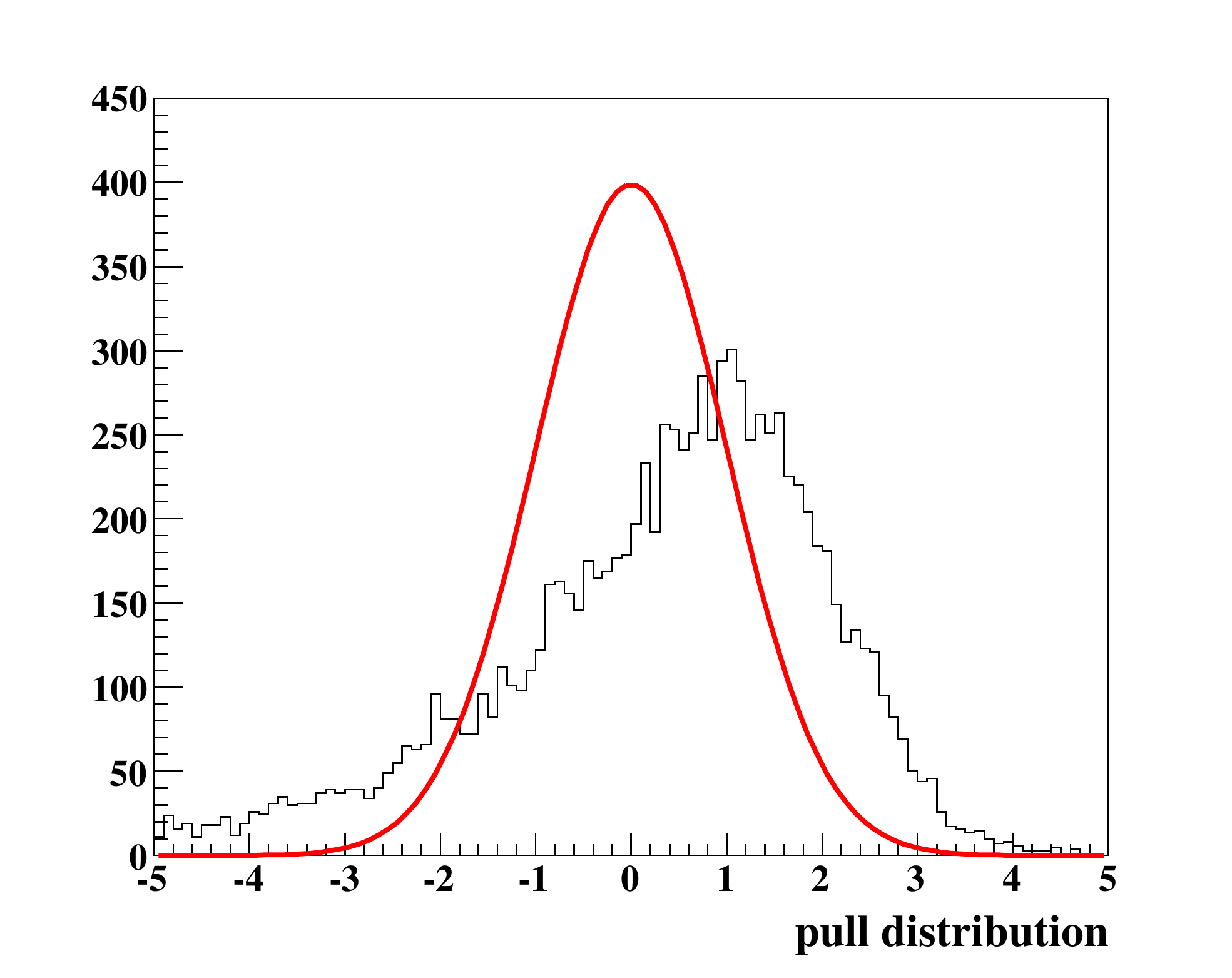}
  }
  \subfigure[]{
  \includegraphics[width=0.3\textwidth]{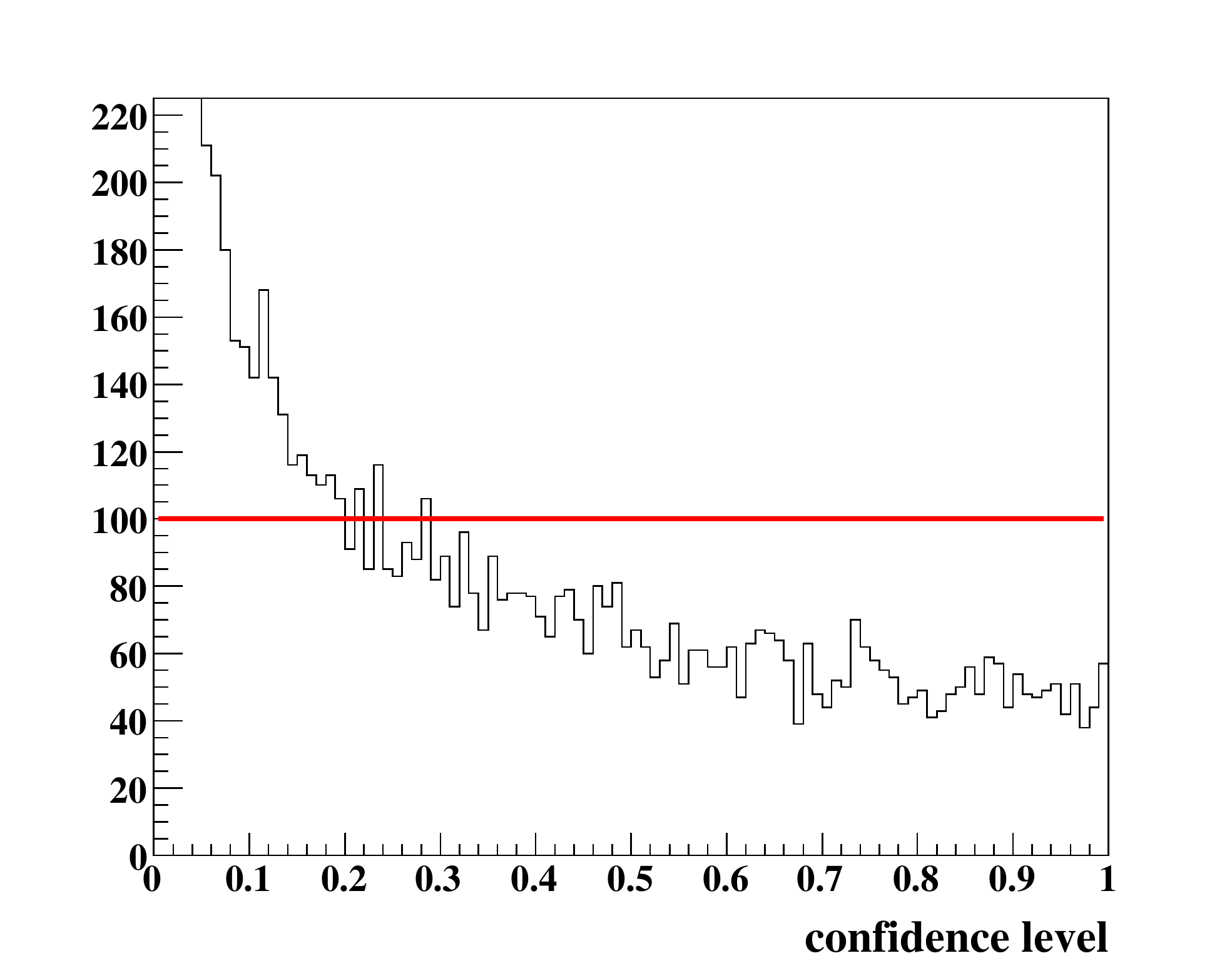}
  }
  \caption[]{\label{fig:chi2-no-off-no-acc-no-bkgd}
    (Color Online) Events generated without background and with 100\% 
    acceptance. The plots represent the
    (a) $\chi^2$ 
    (b) pull and 
    (c) confidence level distributions obtained from a fit to (\ref{eq:schil})
    in which both off-diagonal elements were constrained to be zero. 
    The red lines represent the theoretical distributions, which contain no
    free parameters.
    See text for details.
  }
\end{figure*}

Maximizing the likelihood while
requiring $\rho^0_{1-1} = Re\rho^0_{10} = 0$ yields 
${\rho^0_{00} = 0.651\pm0.009}$. 
Figure~\ref{fig:chi2-no-off-no-acc-no-bkgd} shows the $\chi^2$, pull and 
confidence level distributions obtained for our simulated data set under this
hypothesis. This fit clearly provides a poor description of the data;
the total ${\chi^2/n.d.f. = 36118/(10000-1) = 3.612}$. 

The total $\chi^2/n.d.f.$ values do provide a goodness-of-fit value which 
accurately indicates how well the fit describes the data. Thus, these values
can be used to ascertain the quality of the hypotheses. It is also interesting
to note that the pull and confidence level distributions are also good 
indicators. One could plot these quantities {\em vs.} kinematic
variables to determine where a given hypothesis fails to describe the data.

\subsubsection{\label{section:example:nc}Choosing a Value for $n_c$}

\begin{figure*}
  \centering
  \subfigure[]{
    \includegraphics[width=0.45\textwidth]{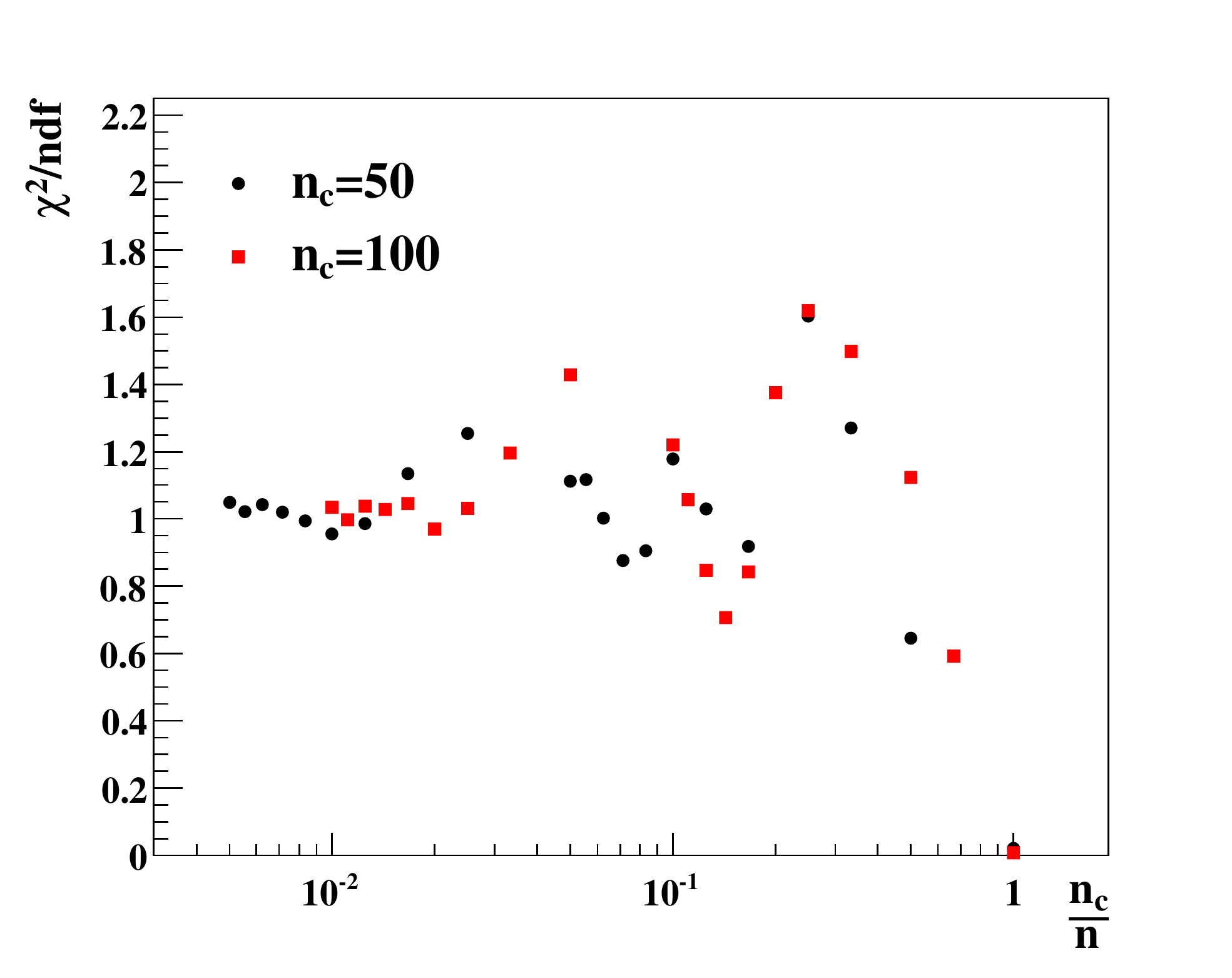}
  }
  \subfigure[]{
    \includegraphics[width=0.45\textwidth]{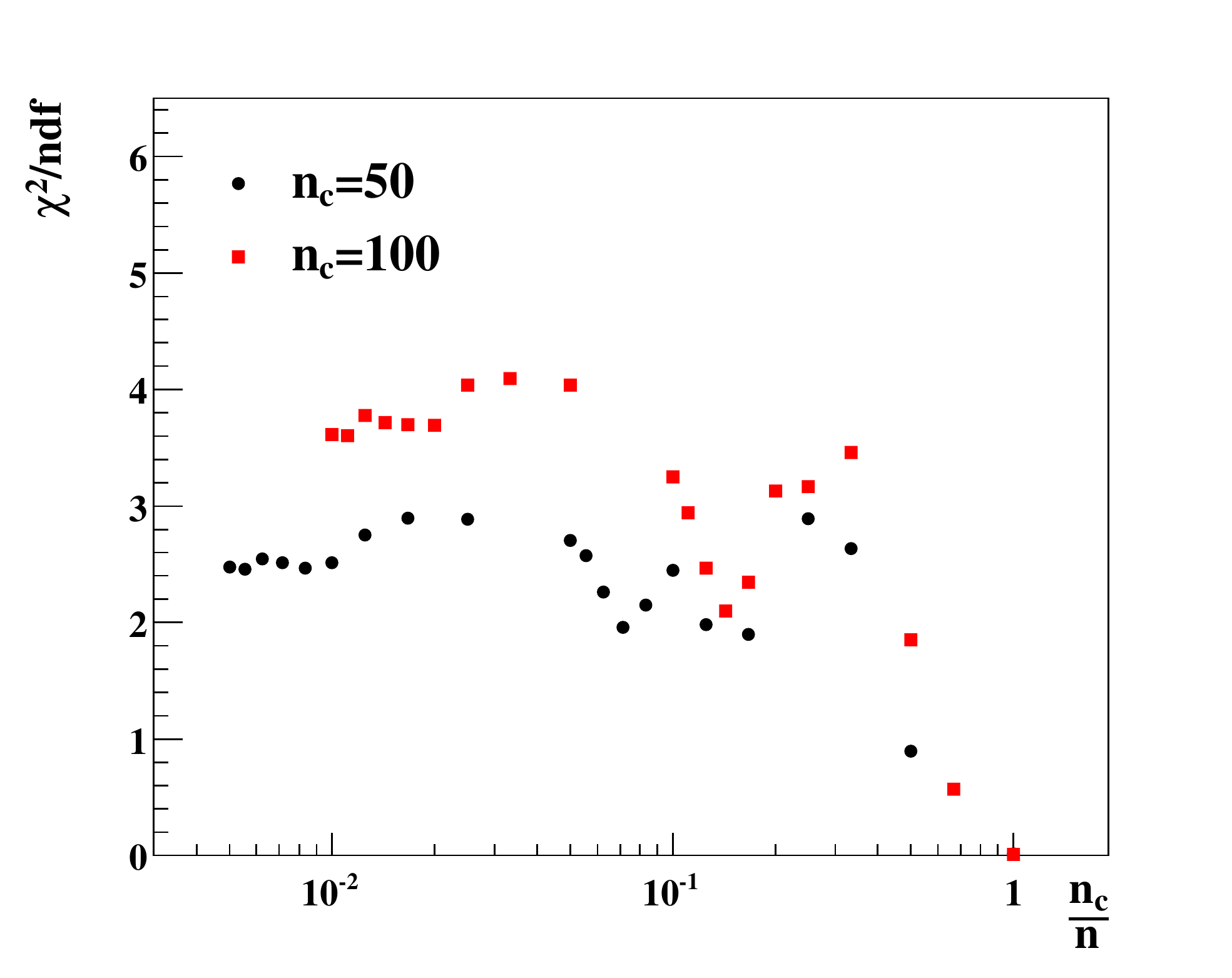}
  }
  \caption[]{\label{fig:nc_ratio}
    (Color Online) Events generated without background and with 100\% 
    acceptance. 
    (a) $\chi^2/n.d.f.$ vs. $\frac{n_c}{n}$ for fits with all three spin 
    density matrix parameters free.
    (b) $\chi^2/n.d.f.$ vs. $\frac{n_c}{n}$ for fits with both off-diagonal
    spin density matrix elements constrained to be zero.
  }
\end{figure*}

In this section, we will examine the value of $n_c$ (the number of
nearest neighbor events used to determine the values of the radii $r_i$). 
Figure~\ref{fig:nc_ratio}(a) shows the $\chi^2/n.d.f$ value obtained for 
two choices of $n_c$ (${n_c = 100}$ and ${n_c = 50}$) for various event sample 
sizes.
The data sets (with sample sizes ranging from 50 to 10,000 events) were all 
generated from the same parent distributions as
in Section~\ref{section:example:simple}. 
Each data set was fit individually to extract the $\rho^0_{MM'}$
values used to obtain the $\chi^2$'s. 

Figure~\ref{fig:nc_ratio}(a) clearly shows that the $\chi^2/n.d.f$ value 
obtained does depend on the choice of $n_c$, if $n_c$ is chosen to be
greater than about 2\% of the total number of events. 
For values of $n_c$ less than 2\% of the event sample size, the $\chi^2/n.d.f.$
is quite stable.
This behavior is expected. If $n_c$ is large relative to $n$, then the 
method is averaging over large fractions of phase space; thus, finer structure 
in the physics will not be properly accounted for.
We also note here that $\chi^2 \rightarrow 0$ as $n_c \rightarrow n$
regardless of the quality of the fit, 
due to how $n_{p}$ is calculated (see (\ref{eq:method-num-predicted})).

Figure~\ref{fig:nc_ratio}(b) shows the value of $\chi^2/n.d.f$ obtained for 
the same two choices of $n_c$ for various event sample sizes for fits to the
data which constrain ${\rho^0_{1-1}=Re\rho^0_{10}=0}$. The values of 
$\chi^2/n.d.f.$ are again stable when $n_c$ is chosen to be less than 2\%
of the data; however, the values depend on the choice of $n_c$.
This behavior is again expected; 
a similar effect can be seen in fits to binned 
data for which the value of $\chi^2/n.d.f$ depends on the number of bins.
The larger the value of $n_c$, the larger the $\chi^2/n.d.f.$ can be. 

Consider a kinematic region where the number density of the data is high, 
leading to a small value of $r_i$. For this case, the largest obtainable value
for $z^2$ is
\begin{equation}
  \lim\limits_{r_i\to 0}z^2_i = \lim\limits_{n_{p_i}\to 0}
  \frac{(n_c - n_{p_i})^2}{n_c + \sigma^2_{p_i}} \approx n_c.
\end{equation}
Thus, there are two competing factors which should be considered when choosing
the value of $n_c$. The ratio of $n_c/n$ must be small enough to permit a true
comparison of the finer structure of the physics; however, the value of 
$n_c$ must be large enough such that the relative statistical uncertainties
do not forbid larger values of $z$. 
Typically, these considerations will dictate $n_c$ be about 1\%-2\% of the 
event sample size. If this results in $n_c$ being less than 50, 
{\em i.e.} if $n < 2500$, then this method will most likely be less effective.
Of course, this statement is not all encompassing; in practice, it will 
depend on the physics.

\begin{figure}[h!]
  \centering
  \includegraphics[width=0.45\textwidth]{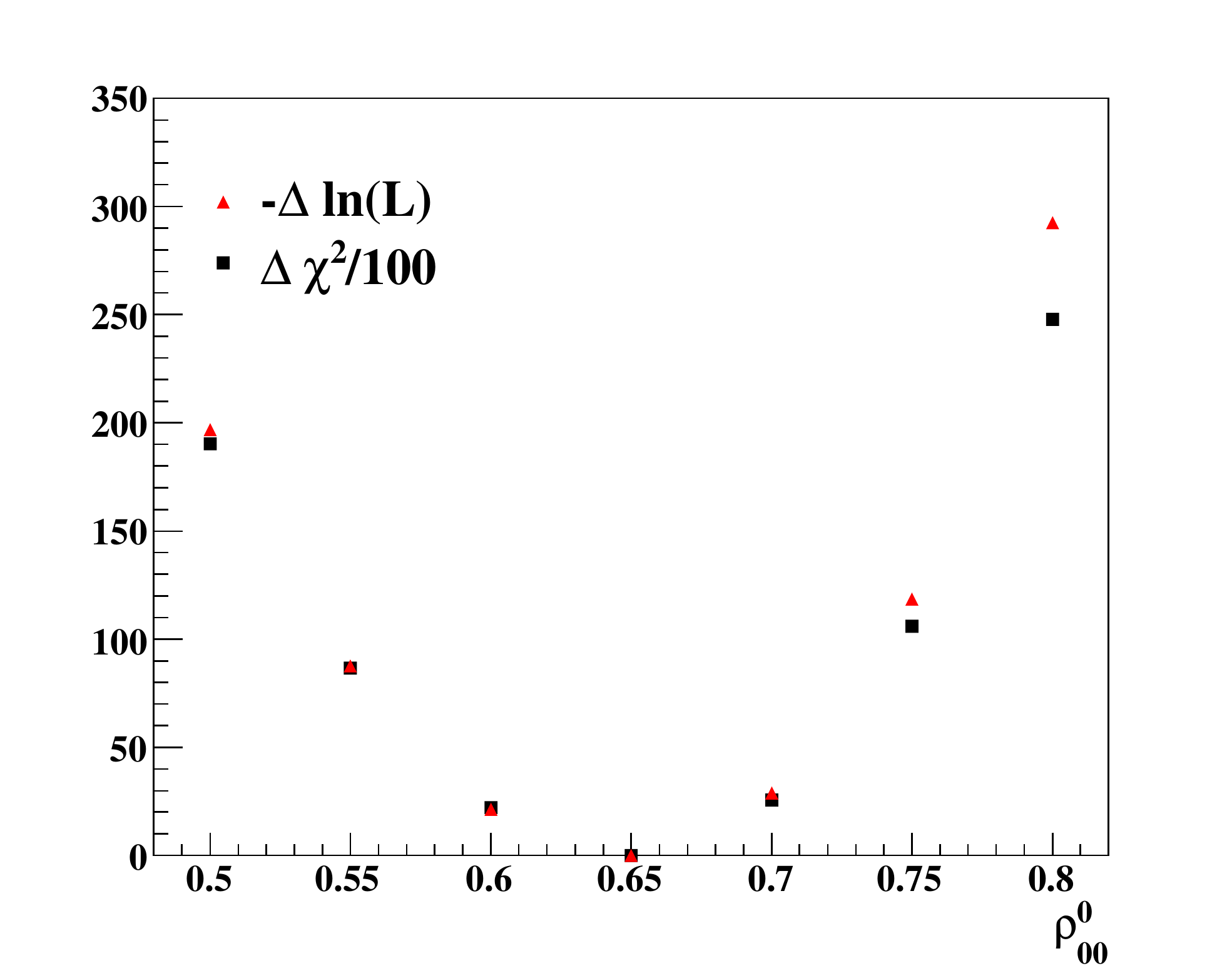}
  \caption[]{\label{fig:diff-comp}
    (Color Online) Events generated without background and with 100\% 
    acceptance. The quantities plotted are $-\Delta\ln{\mathcal{L}}$ vs.
    $\rho^0_{00}$ (red triangles) and $\Delta\chi^2/100$ vs.
    $\rho^0_{00}$ (black squares).    
  }
\end{figure}
\subsubsection{$\Delta \chi^2$ vs. $-\Delta \log{\mathcal{L}}$}
It is also useful to examine how changes in the likelihood map onto changes 
in the $\chi^2$ values obtained using this method. Here we trace out both
goodness-of-fit quantities near the optimal value of $\rho^0_{00}$.
Figure~\ref{fig:diff-comp} shows the changes in $-\ln{\mathcal{L}}$ and
$\chi^2$ {\em vs} $\rho^0_{00}$ for the data set used in 
Section~\ref{section:example:simple}
($n = 10,000$ and $n_c = 100$). Clearly, an increase in 
$-\ln{\mathcal{L}}$ does corresponds to an increase in $\chi^2$

\subsection{\label{section:example:acc}Including Detector Acceptance}
\begin{figure}[h!]
  \centering
  \includegraphics[width=0.45\textwidth]{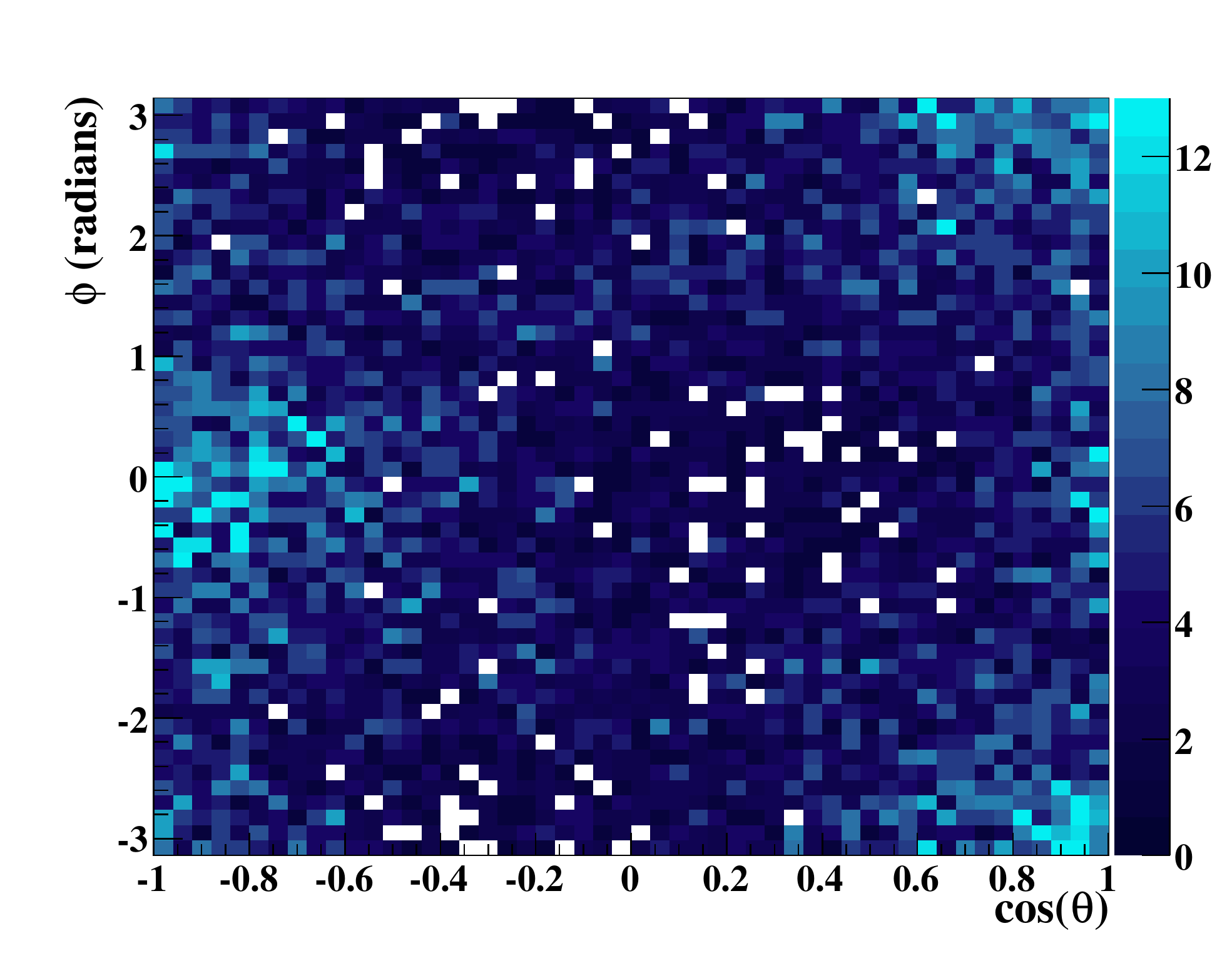}
  \caption[]{\label{fig:gen-acc-no-bkgd}
    (Color Online) 
    $\phi$ (radians) vs $\cos{\theta}$:
    Events generated without background and with detector 
    acceptance given by (\ref{eq:detector-acc}). 
  }
\end{figure}

We will now extend our example by including detector acceptance. 
The physics used in this section will be identical to that of 
Section~\ref{section:example:simple};
however, we will now include a detector efficiency, $\eta$, given by
\begin{equation}
  \label{eq:detector-acc}
 \eta(\theta,\phi) = \frac{1}{2}(2 - |\cos{\theta}\sin{\phi}|).
\end{equation}
We again generated 10,000 data events and 100,000 Monte Carlo events from the
same parent distributions as in Section~\ref{section:example:simple} 
convolved with the detector 
acceptance given in (\ref{eq:detector-acc}).
Figure~\ref{fig:gen-acc-no-bkgd} shows the decay angular distribution of this
data set. The effects of the detector acceptance are clearly visible when 
compared with Figure~\ref{fig:gen-no-acc-no-bkgd}.

\begin{figure*}
  \centering
  \subfigure[]{
  \includegraphics[width=0.3\textwidth]{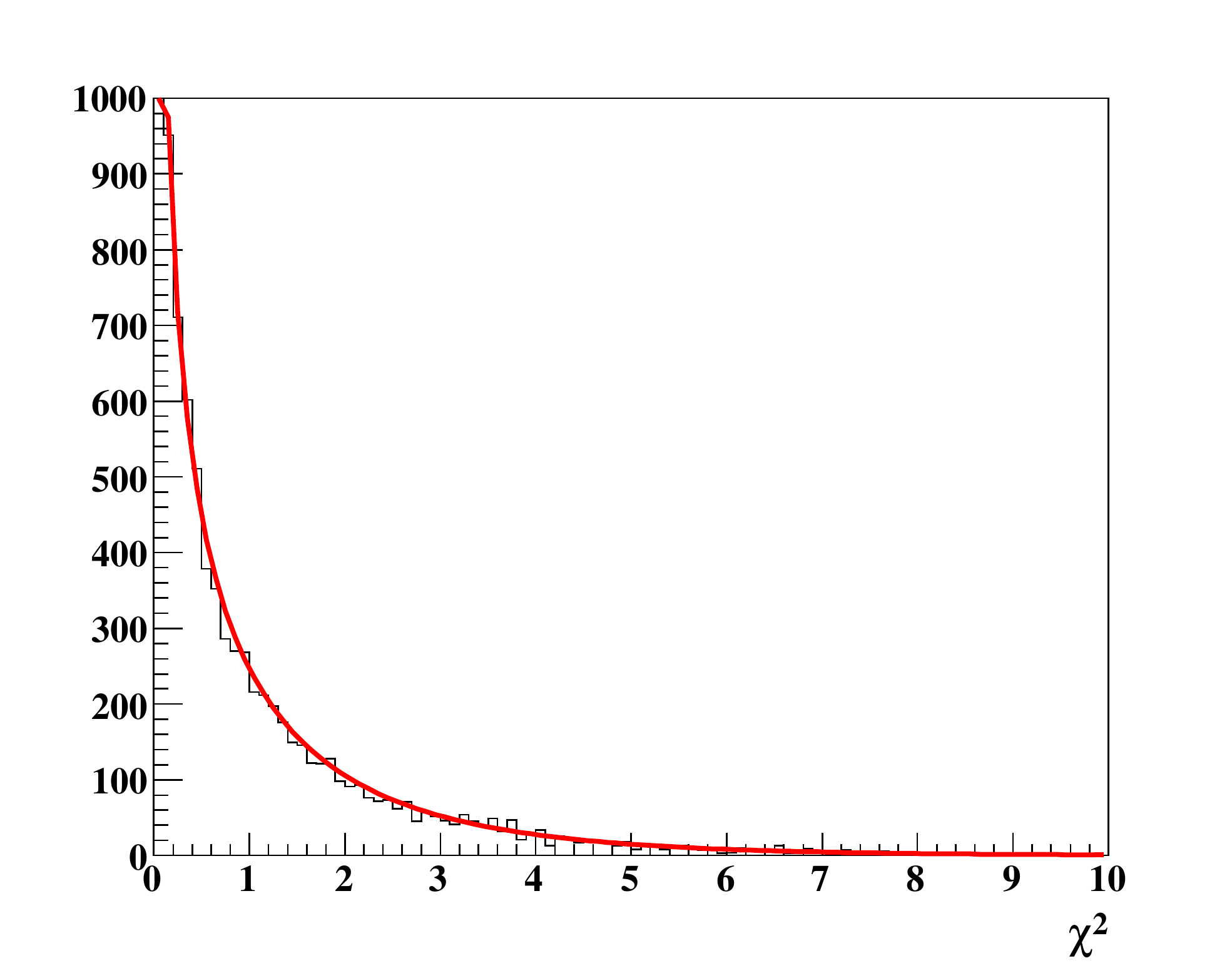}
  }
  \subfigure[]{
  \includegraphics[width=0.3\textwidth]{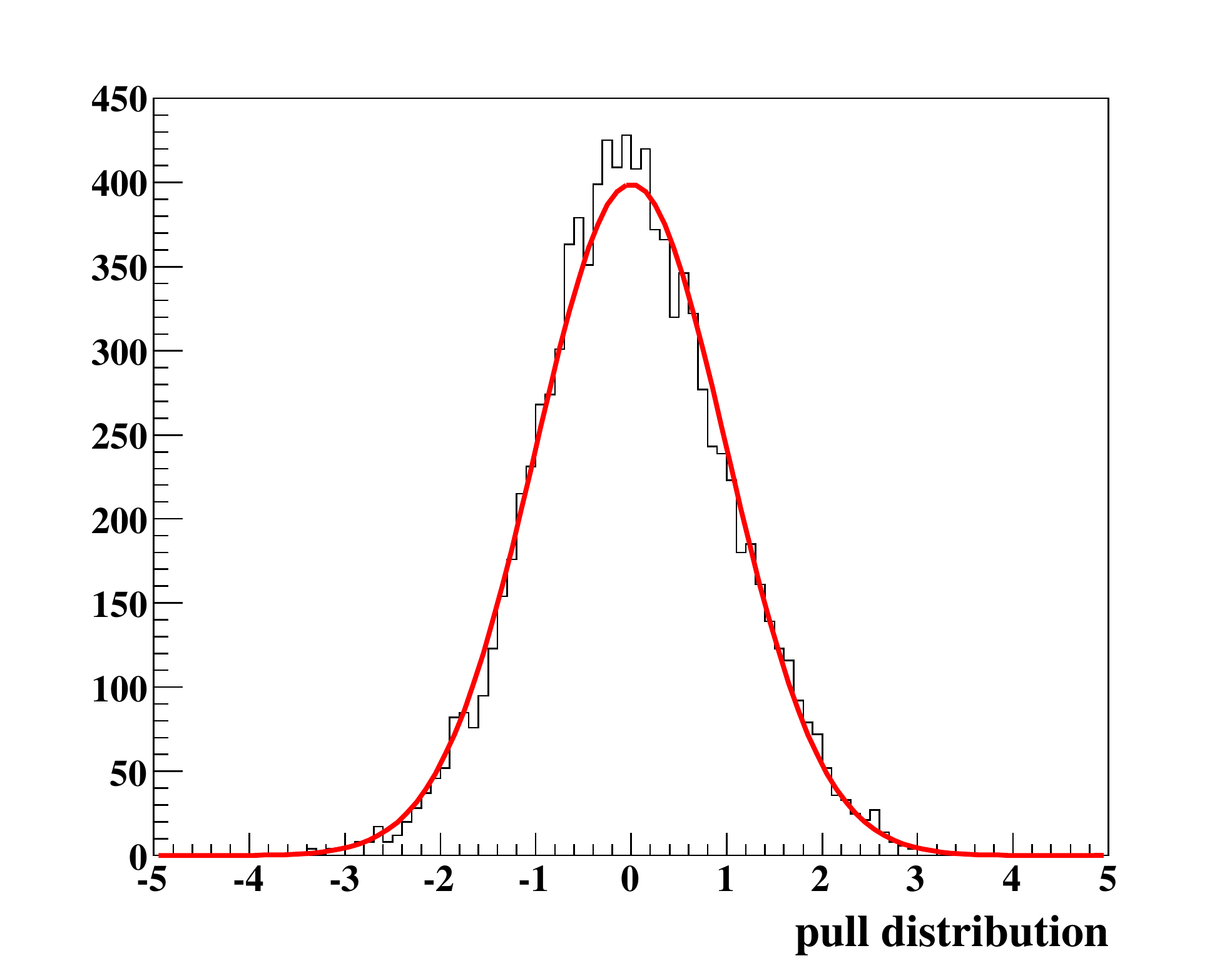}
  }
  \subfigure[]{
  \includegraphics[width=0.3\textwidth]{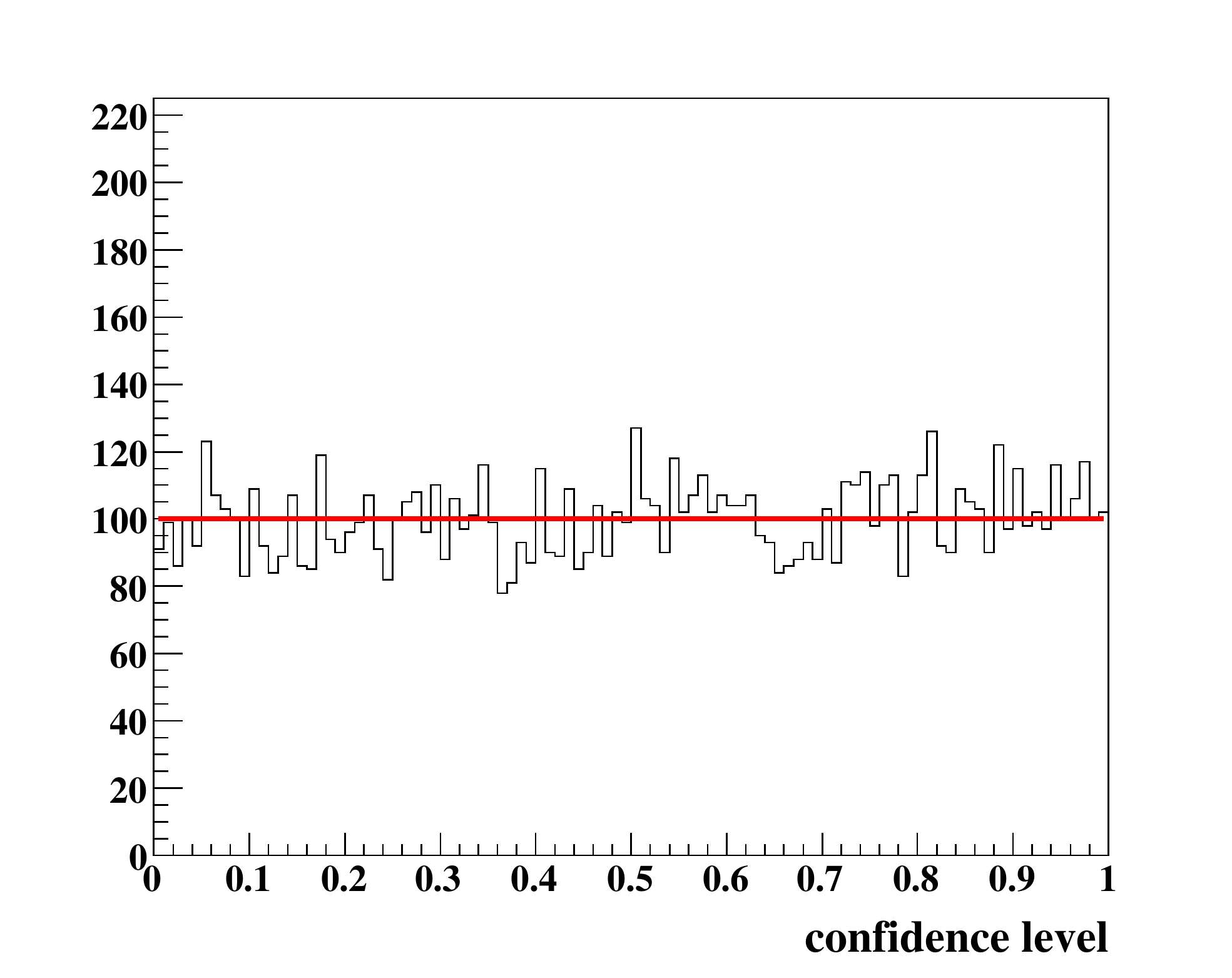}
  }
  \caption[]{\label{fig:chi2-acc-no-bkgd}
    (Color Online) Events generated without background and with detector 
    acceptance given by (\ref{eq:detector-acc}).  
    (a) $\chi^2$ 
    (b) pull and
    (c) confidence level distributions.
    The red lines represent the theoretical distributions, which contain no
    free parameters.
  }
\end{figure*}

An unbinned maximum likelihood fit was performed and the following values were
obtained for $\rho^0_{00}$:
\begin{subequations}
  \label{eq:rho-acc-no-bkgd}
  \begin{equation}
  \rho^0_{00} = 0.634\pm0.009
  \end{equation}
  \begin{equation}  
  \rho^0_{1-1} = 0.055\pm0.005 
  \end{equation}
  \begin{equation}
    Re\rho^0_{10} = 0.096\pm0.004.
  \end{equation}
\end{subequations}
To obtain $\chi^2$ values, we apply the procedure just as in 
Section~\ref{section:example:simple}. 
Figure~\ref{fig:chi2-acc-no-bkgd} shows the $\chi^2$, pull and confidence level
distributions obtained for this simulated data set. The values are again in 
excellent agreement with the theoretical distributions. The total 
$\chi^2/n.d.f. = 9715.65/(10000-3) = 0.972$, indicating that this fit does
provide an excellent description of the data (as it should).

\subsection{\label{section:example:bkgd}Including Background Events}
In many physics analyses, there is a sample of non-interfering background 
events which can not be separated from the signal. We will now deal with this
situation. The signal sample will simply be that of 
Section~\ref{section:example:acc}; 
thus, we will be including detector acceptance effects as well.
For the background, we chose to generate it according to 3-body phase space 
weighted by a linear function in $m_{3\pi}$ and 
\begin{equation}
  W_b(\theta,\phi) = \frac{1}{6\pi}\left(1 + |\sin{\theta}\cos{\phi}|\right)
\end{equation}
in the decay angles. The number of background events generated, $n_b$, was 
10,000 (the background events were also subjected to the detector acceptance
given in (\ref{eq:detector-acc})).
Figure~\ref{fig:decay-bkgd-acc}(a) shows the 
$\pi^+\pi^-\pi^0$ mass spectrum for all generated events and for just the 
background. The generated
decay angular distributions for all events, along with only the signal and
background are shown in Figures~\ref{fig:decay-bkgd-acc}(b), (c) and (d). 
There is clearly no
way to separate out the signal events through the use of a cut.

\subsubsection{\label{section:example:q-factors}Signal Extraction}

The method used to extract the signal events is described in detail 
in~\cite{cite:sig-bkgd}. 
It accurately preserves all kinematic correlations in the data
while separating signal and background events. 
Each event is assigned a signal weight factor, $Q$, or equivalently, a 
background weight factor, $1-Q$. 
These $Q$-factors are then used to weight each
event's contribution to the ``log likelihood'' during unbinned maximum 
likelihood fits to extract physical observables. 

The method works in a very similar way to the one presented in this paper.
A metric is defined in the space of all relevant kinematic variables and the
$n_c$ nearest neighbor events (we chose $n_c = 100$) are selected. 
For this example, the metric defined in (\ref{eq:example-metric}) 
can again be used. 
Since each subset of events occupies a very small region of phase space, the
$m_{3\pi}$ distributions can be used to determine each event's $Q$-factor
while preserving the correlations present in the remaining kinematic variables
($\cos{\theta}$ and $\phi$).

To this end, unbinned maximum likelihood fits were carried out for each event,
using its nearest neighbors, to determine the parameters 
${\vec{\alpha}= (b_0,b_1,s,\sigma)}$ in the p.d.f.
\begin{equation}
  F(m_{3\pi},\vec{\alpha}) 
  = \frac{B(m_{3\pi},\vec{\alpha}) + S(m_{3\pi},\vec{\alpha})}
  {\int \left(B(m_{3\pi},\vec{\alpha}) + S(m_{3\pi},\vec{\alpha})\right)
    dm_{3\pi}},
\end{equation}
where,
\begin{equation}
  S(m_{3\pi},\vec{\alpha}) 
  = s\cdot V(m_{3\pi},m_{\omega},\Gamma_{\omega},\sigma)
\end{equation}
parameterizes the signal as a Voigtian 
(convolution of a Breit-Wigner and a Gaussian)
with mass ${m_{\omega}=0.78256}$~GeV/c$^2$, 
natural width ${\Gamma_{\omega}=0.00844}$~GeV/c$^2$ and resolution
$\sigma$. The parameter $s$ sets the overall strength of the signal.
The background, in each small phase space region, was parameterized by the
linear function
\begin{equation}
  B(m_{3\pi},\vec{\alpha}) = b_1 m_{3\pi} + b_0.
\end{equation}
The $Q$-factor for the event was then calculated as
\begin{equation}
  Q_i = \frac{S(m_{3\pi}^i,\hat{\alpha}_i)}
  {S(m_{3\pi}^i,\hat{\alpha}_i) + B(m_{3\pi}^i,\hat{\alpha}_i)},
\end{equation}
where $m_{3\pi}^i$ is the event's $3\pi$ mass and $\hat{\alpha}_i$ are the
estimators for the parameters obtained from the $i^{th}$ event's fit. 

\begin{figure*}
  \begin{center}
    \subfigure[]{
      \includegraphics[width=0.45\textwidth]{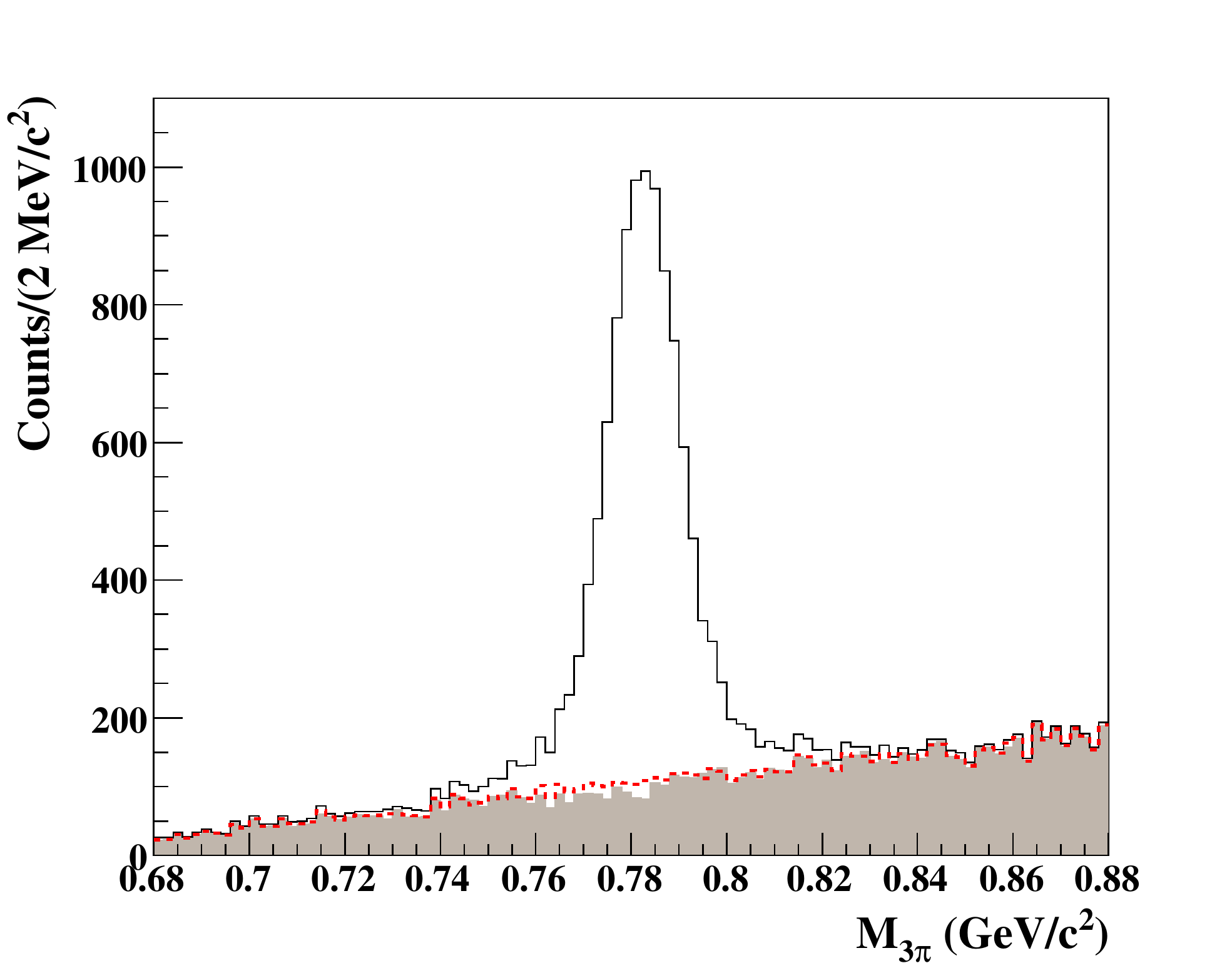}
    }
    \subfigure[]{
      \includegraphics[width=0.45\textwidth]{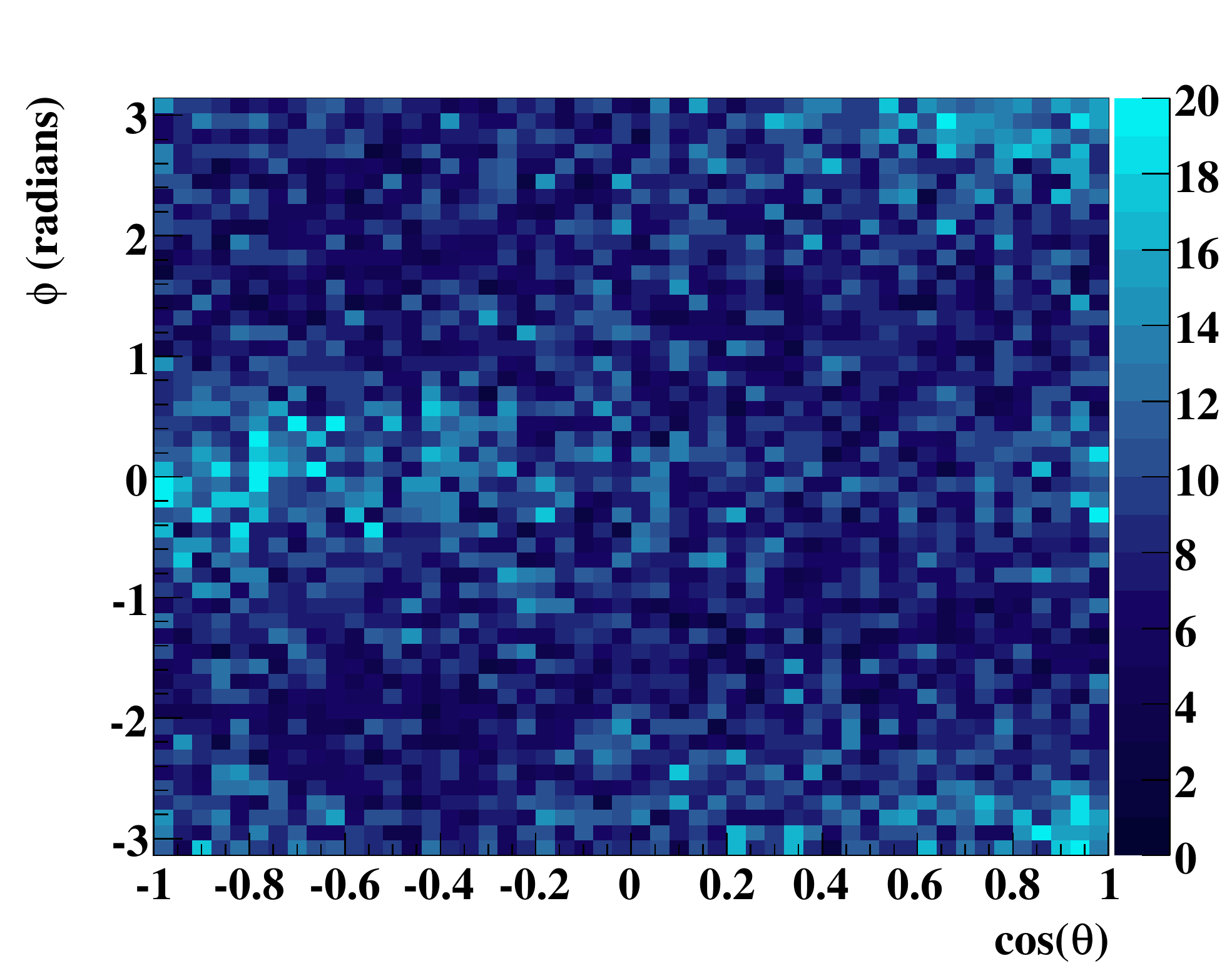}
    }
    \subfigure[]{
     \includegraphics[width=0.45\textwidth]{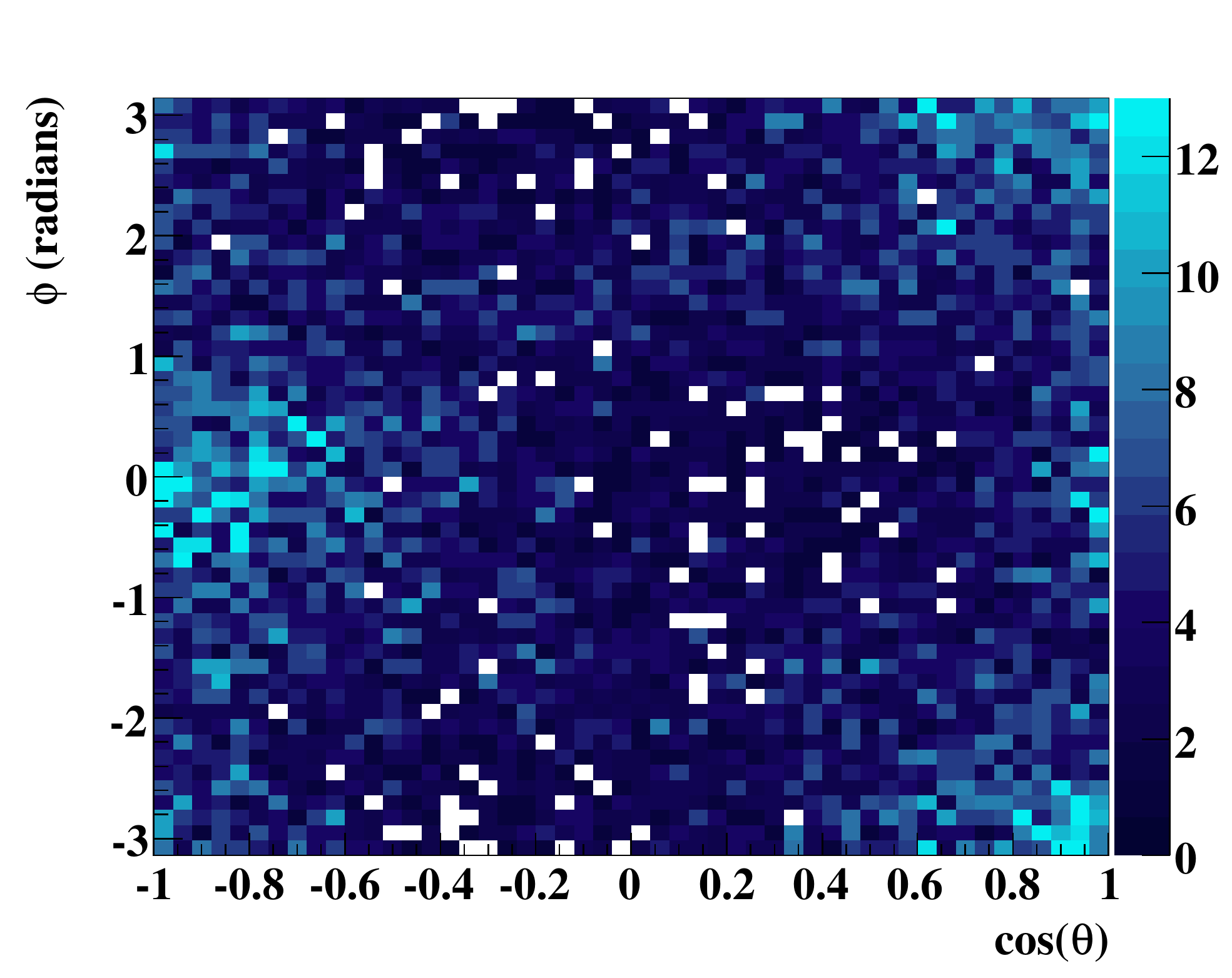}
    }
    \subfigure[]{
      \includegraphics[width=0.45\textwidth]{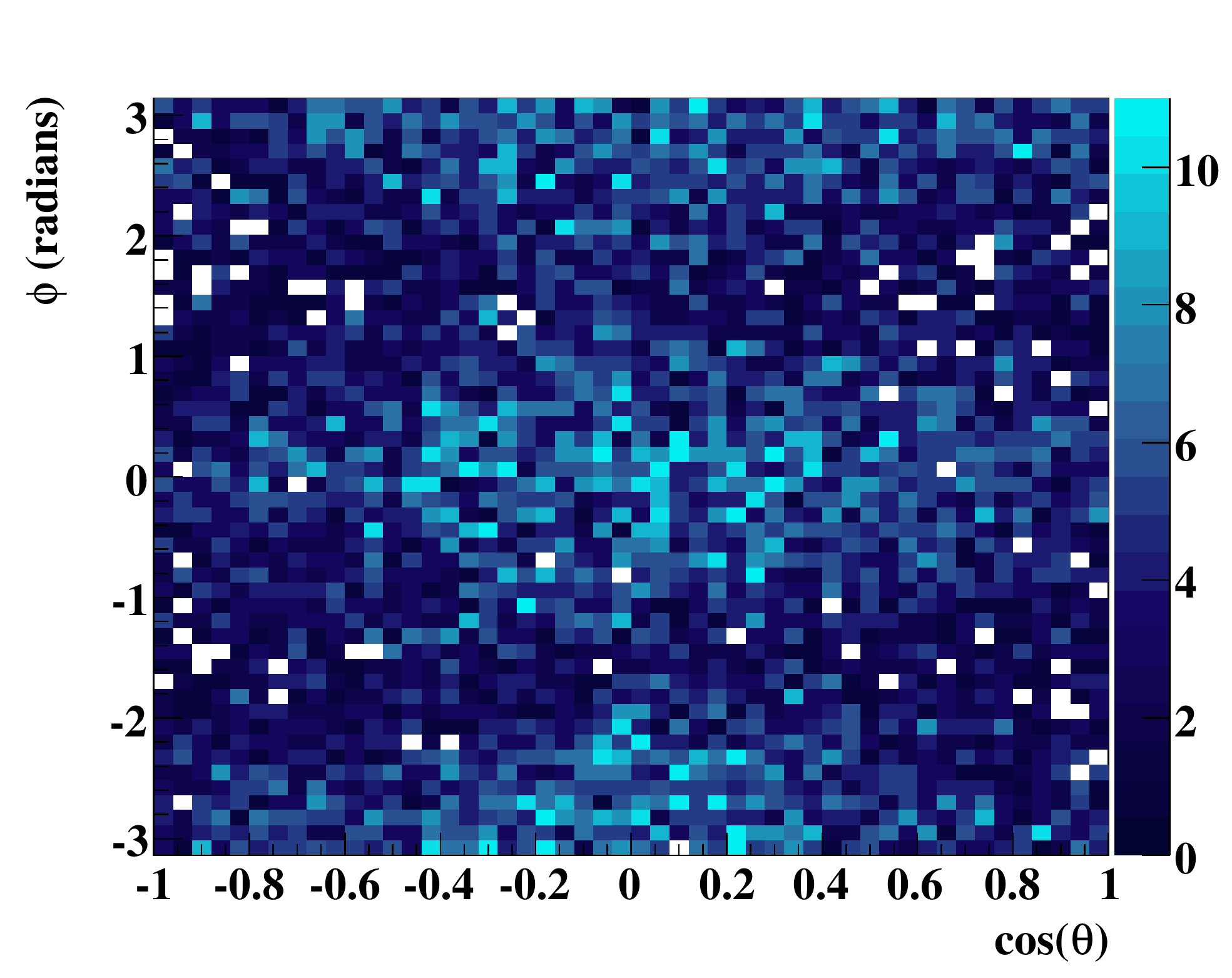}
    }
   \subfigure[]{
     \includegraphics[width=0.45\textwidth]{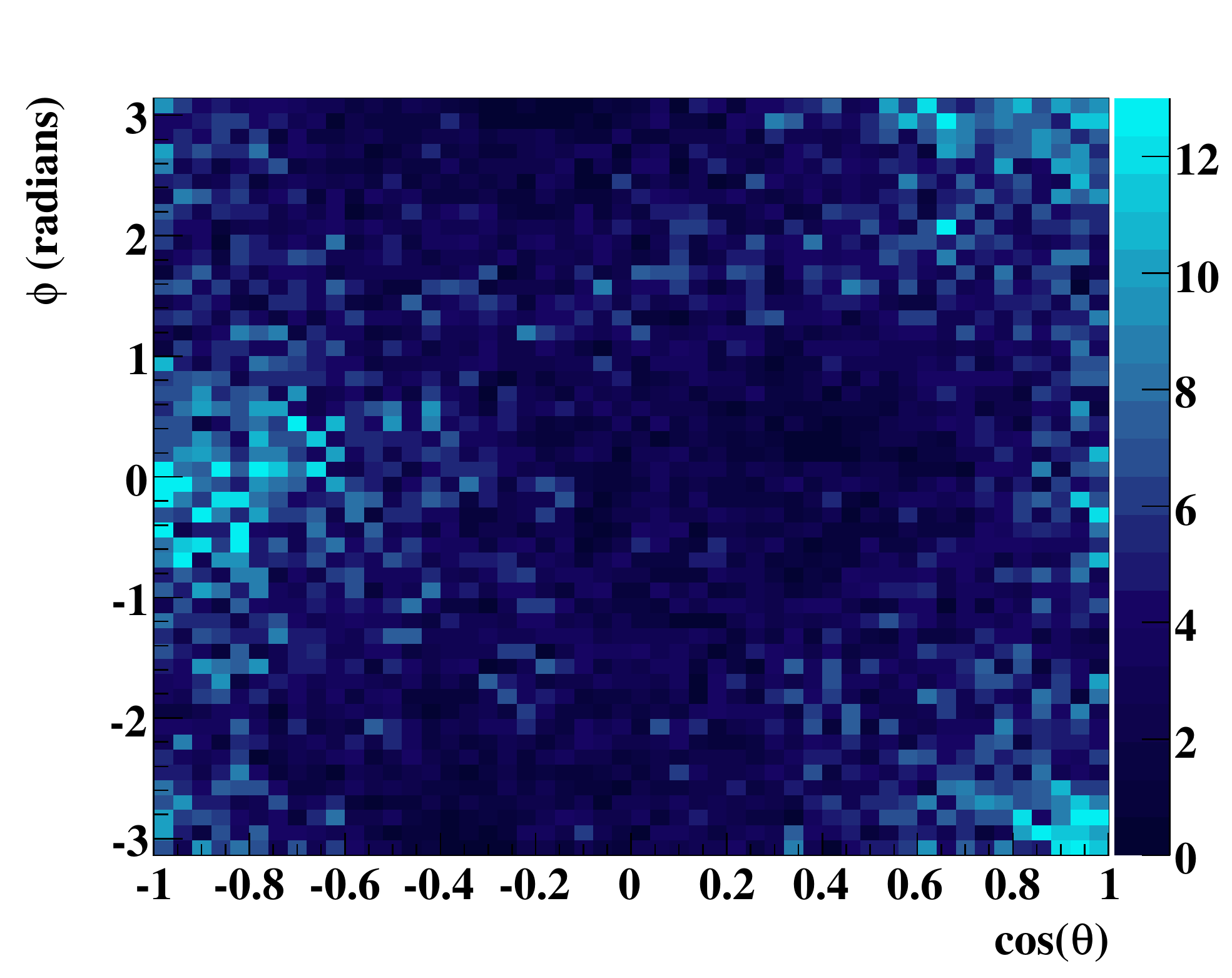}
    }
    \subfigure[]{
      \includegraphics[width=0.45\textwidth]{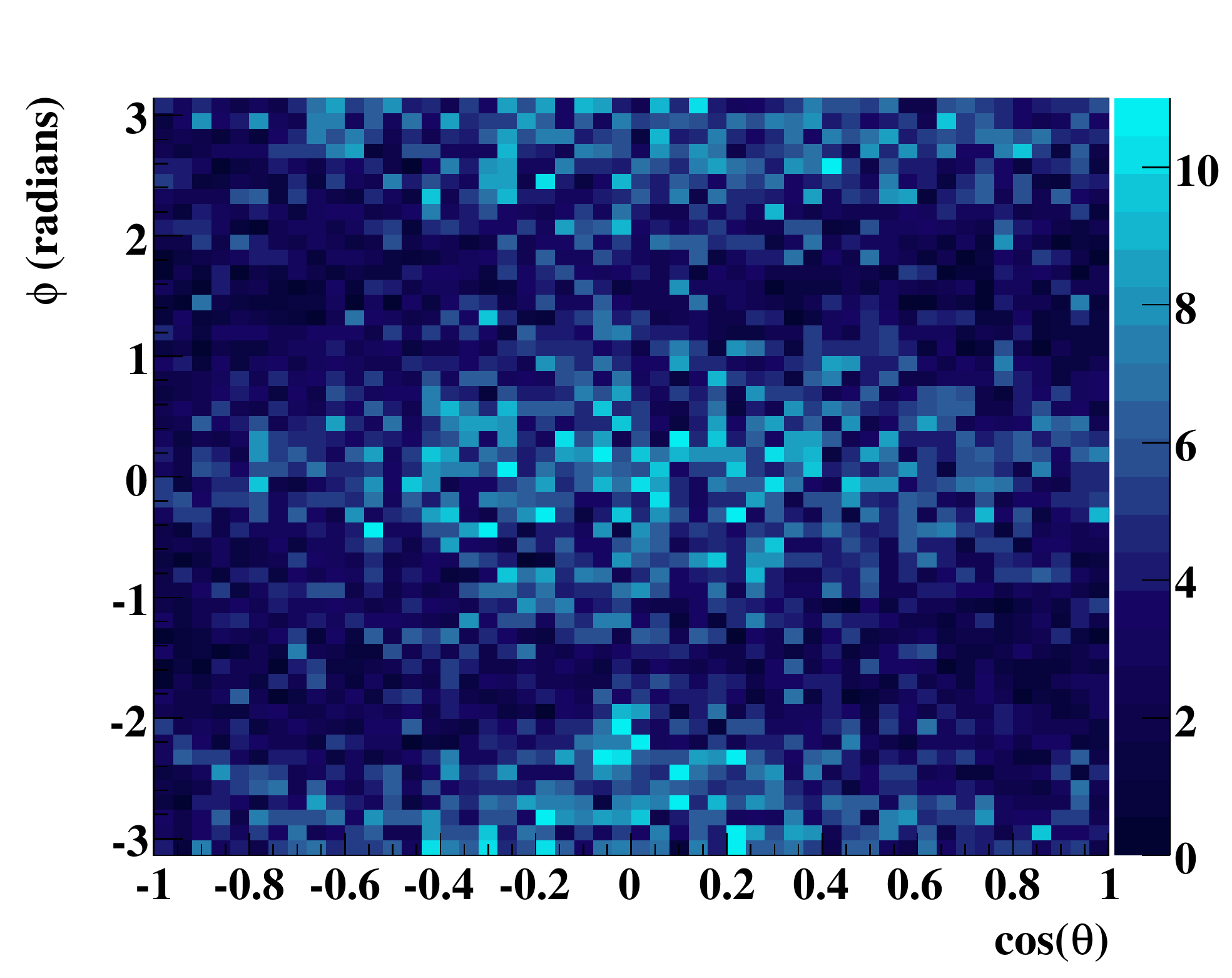}
    }
  \caption[]{\label{fig:decay-bkgd-acc}
    (Color Online)
    (a) The $m_{3\pi}$ distributions for all generated events (unshaded), only
    background events (shaded) and for all events weighted by the background
    factors, $1-Q$ (dashed red).
    (b) $\phi$ vs. $\cos{\theta}$ for all generated events.
    (c) $\phi$ vs. $\cos{\theta}$ for only signal events.
    (d) $\phi$ vs. $\cos{\theta}$ for only background events.
    (e) $\phi$ vs. $\cos{\theta}$ for all events weighted by signal factors, 
    $Q$.
    (f) $\phi$ vs. $\cos{\theta}$ for all events weighted by background 
    factors, $1-Q$.
  }
  \end{center}
\end{figure*}

Figures~\ref{fig:decay-bkgd-acc}(e) and (f) show the extracted signal and
background distributions, {\em i.e.} they show the events weighted
by $Q$ and $1-Q$ respectively. The agreement with the generated distributions
(see Figures~\ref{fig:decay-bkgd-acc}(c) and (d)) is very good.
We conclude this section by noting that
the full covariance matrix obtained from each event's fit, $C_{\alpha}$, can
be used to calculate the uncertainty in $Q$ as
\begin{equation}
  \label{eq:sigma-q}
  \sigma^2_Q = \sum \limits_{ij} \frac{\partial Q}{\partial \alpha_i}
  (C_{\alpha}^{-1})_{ij} \frac{\partial Q}{\partial \alpha_j}.
\end{equation}

\subsubsection{Extracting Physical Observables}

The likelihood function used to obtain the spin density matrix elements, 
defined in (\ref{eq:log-likelihood}), 
can be modified to account for the presence of background events as follows:
\begin{equation}
  \label{eq:log-L-Q}
  -\ln{\mathcal{L}} = -\sum\limits_i^{n + n_b} Q_i \ln{W(\theta_i,\phi_i)}.
\end{equation}
Thus, the $Q$-factors are used to weight each event's contribution to the 
likelihood.
Using the $Q$-factors obtained in Section~\ref{section:example:q-factors}, 
minimizing (\ref{eq:log-L-Q}) yields
\begin{subequations}
  \begin{equation}
  \rho^0_{00} = 0.640\pm0.009
  \end{equation}
  \begin{equation}  
  \rho^0_{1-1} = 0.051\pm0.005 
  \end{equation}
  \begin{equation}
    Re\rho^0_{10} = 0.095\pm0.004,
  \end{equation}
\end{subequations}
which are in good agreement with the generated values.

\subsubsection{Applying the Method}

The metric used to obtain the $n_c$ nearest neighbor events is again 
(\ref{eq:example-metric});
however, the number of measured events contained within each hypersphere is now
given by
\begin{equation}
  n_{m_i} = \sum\limits_i^{n_c} Q_i,
\end{equation}
with uncertainty
\begin{equation}
  \label{eq:n_m-error}
  \sigma^2_{m_i} = n_{m_i} 
  + \sum\limits_{j,k} \sigma_{Q_i}^j \rho^{jk} \sigma_{Q_i}^k,
\end{equation}
where the sums ($j,k$) are over the $n_c$ events used to calculate the $i^{th}$
event's $Q$-factor and
$\rho^{jk}$ is the correlation factor between events $j$ and $k$. This 
factor is equal to the fraction of shared nearest neighbor events used in 
calculating the $Q$-factors for these events. Thus, (\ref{eq:n_m-error}) 
accounts for
both the statistical uncertainty in the signal yield, along with the 
uncertainty in the signal-background separation, {\em i.e.} the uncertainty
in the $Q$-factors.

\begin{figure*}
  \centering
  \subfigure[]{
  \includegraphics[width=0.3\textwidth]{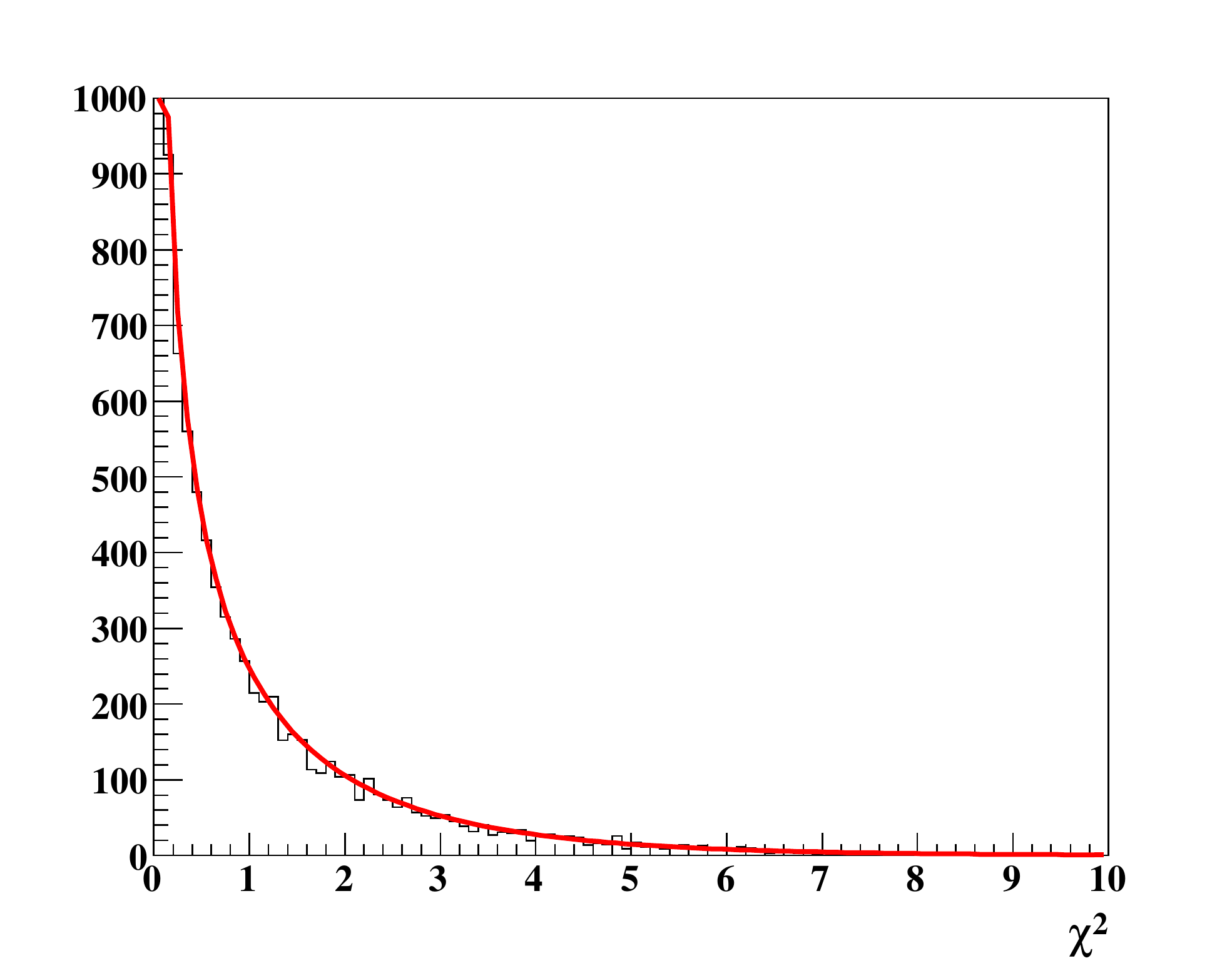}
  }
  \subfigure[]{
  \includegraphics[width=0.3\textwidth]{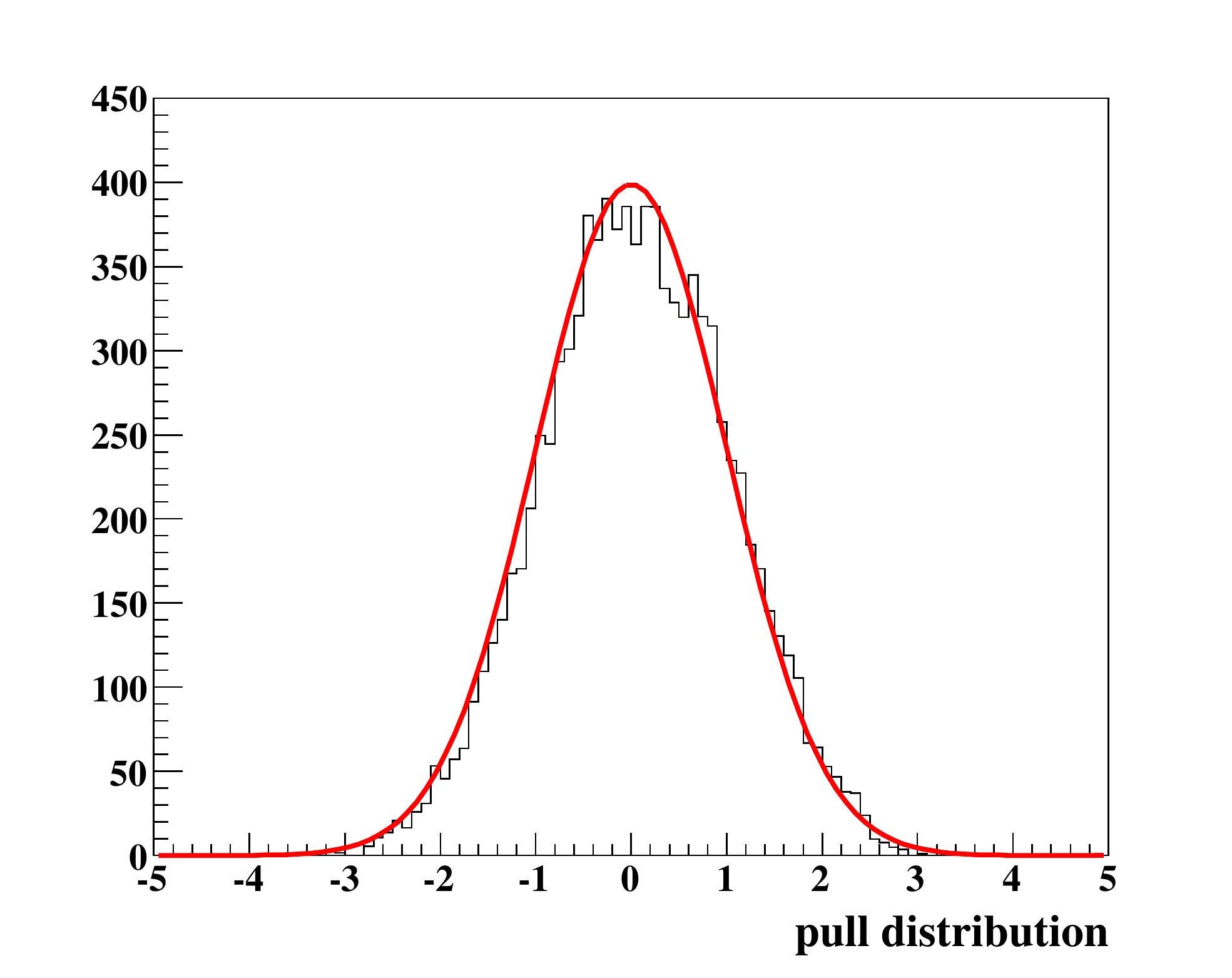}
  }
  \subfigure[]{
  \includegraphics[width=0.3\textwidth]{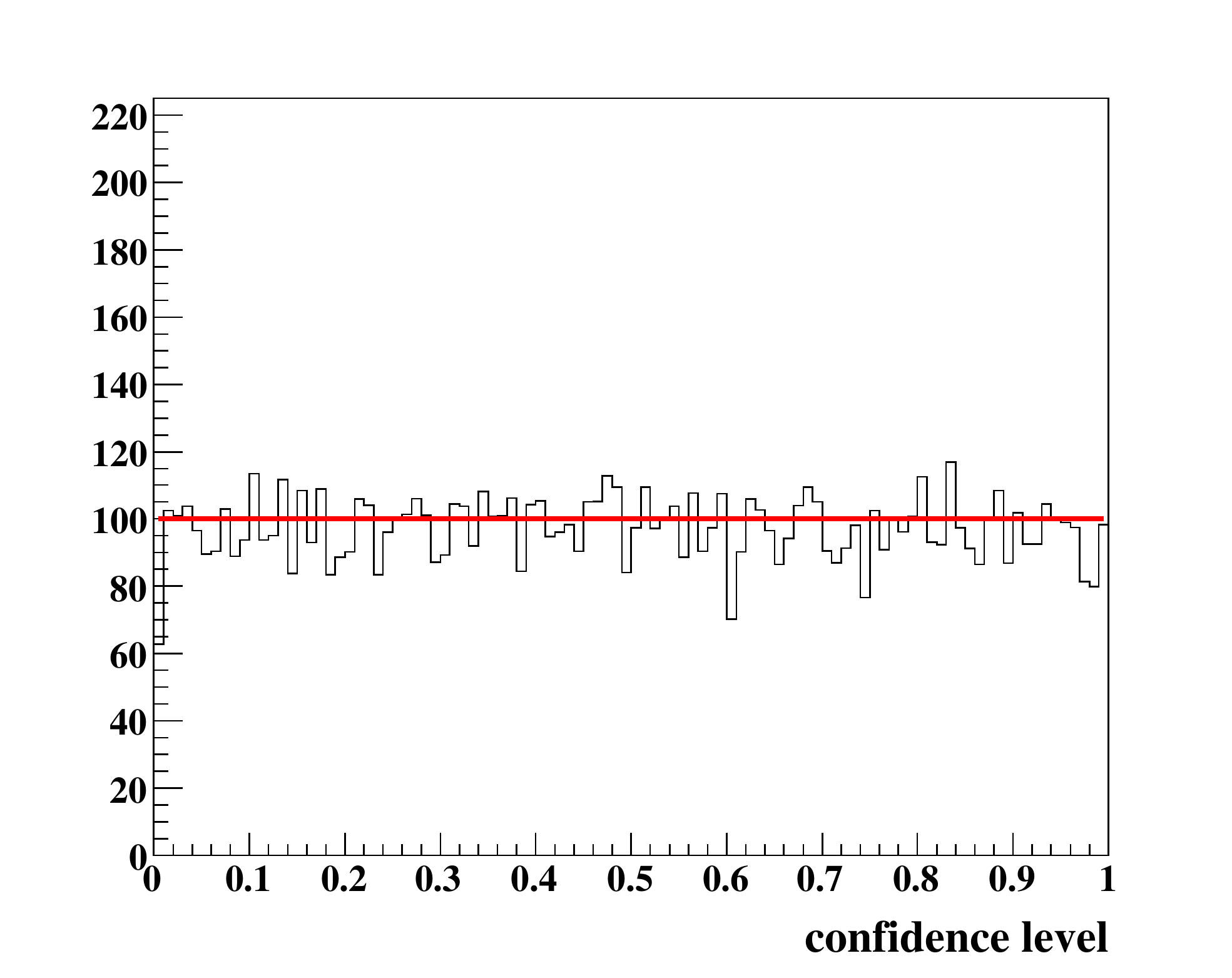}
  }
  \caption[]{\label{fig:chi2-acc-bkgd}
    (Color Online) Events generated with background and with detector 
    acceptance given by (\ref{eq:detector-acc}).  
    (a) $\chi^2$ 
    (b) pull and 
    (c) confidence level distributions.
    The red lines represent the theoretical distributions, which contain no
    free parameters.
  }
\end{figure*}

The number of predicted events, along with its uncertainty, is calculated the 
same way as in Section~\ref{section:example:simple} 
and the $z$ values are again obtained using (\ref{eq:chi2-event}).
Figure~\ref{fig:chi2-acc-bkgd} shows the $\chi^2$, pull and confidence level
distributions obtained for our simulated data set. The values are again in
excellent agreement with the theoretical distributions. This indicates that
we have properly handled the uncertainties in the signal.
The total ${\chi^2/n.d.f. = 9496.75/(9714.82-3)= 0.978}$.

\begin{figure*}
  \centering
  \subfigure[]{
  \includegraphics[width=0.3\textwidth]{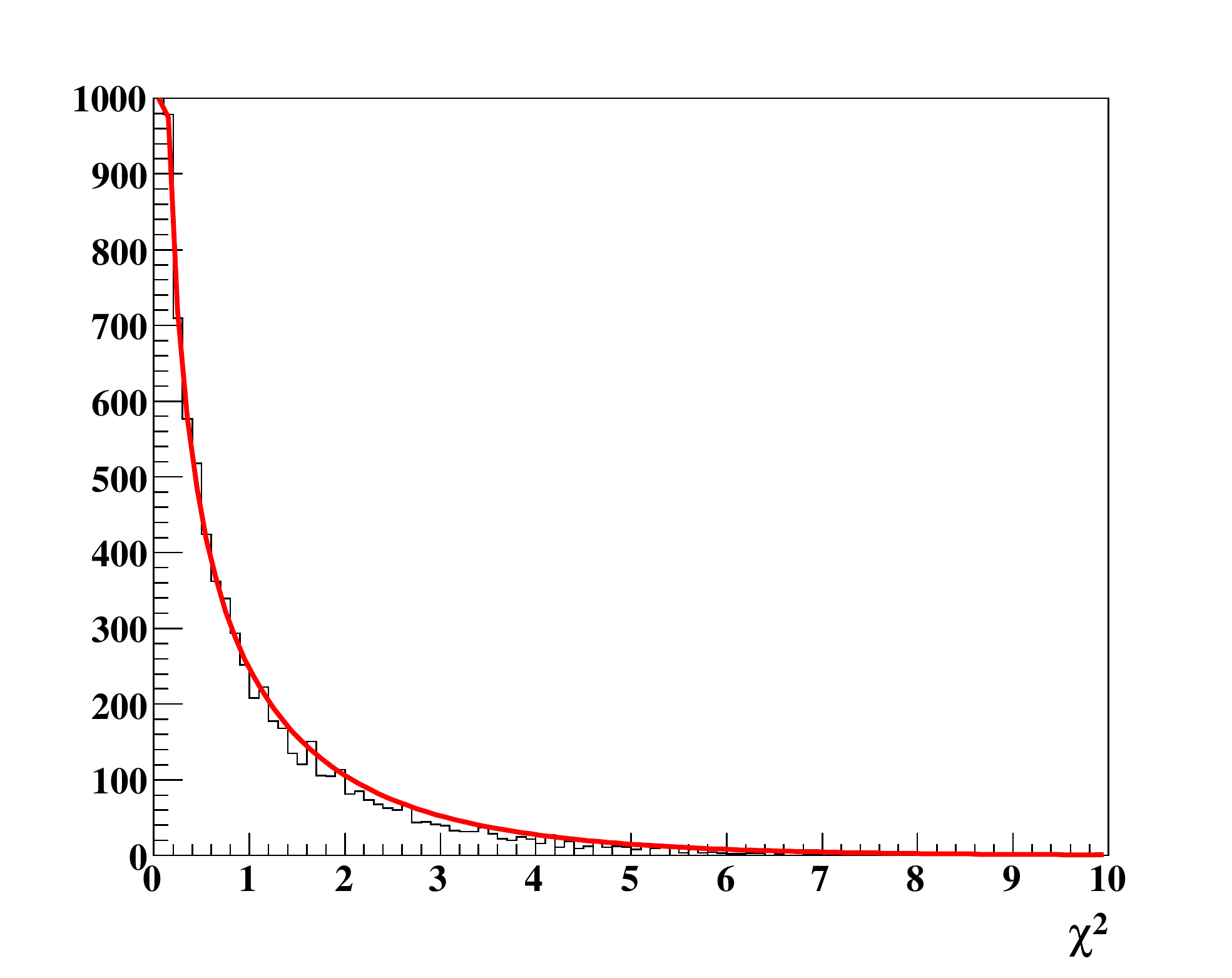}
  }
  \subfigure[]{
  \includegraphics[width=0.3\textwidth]{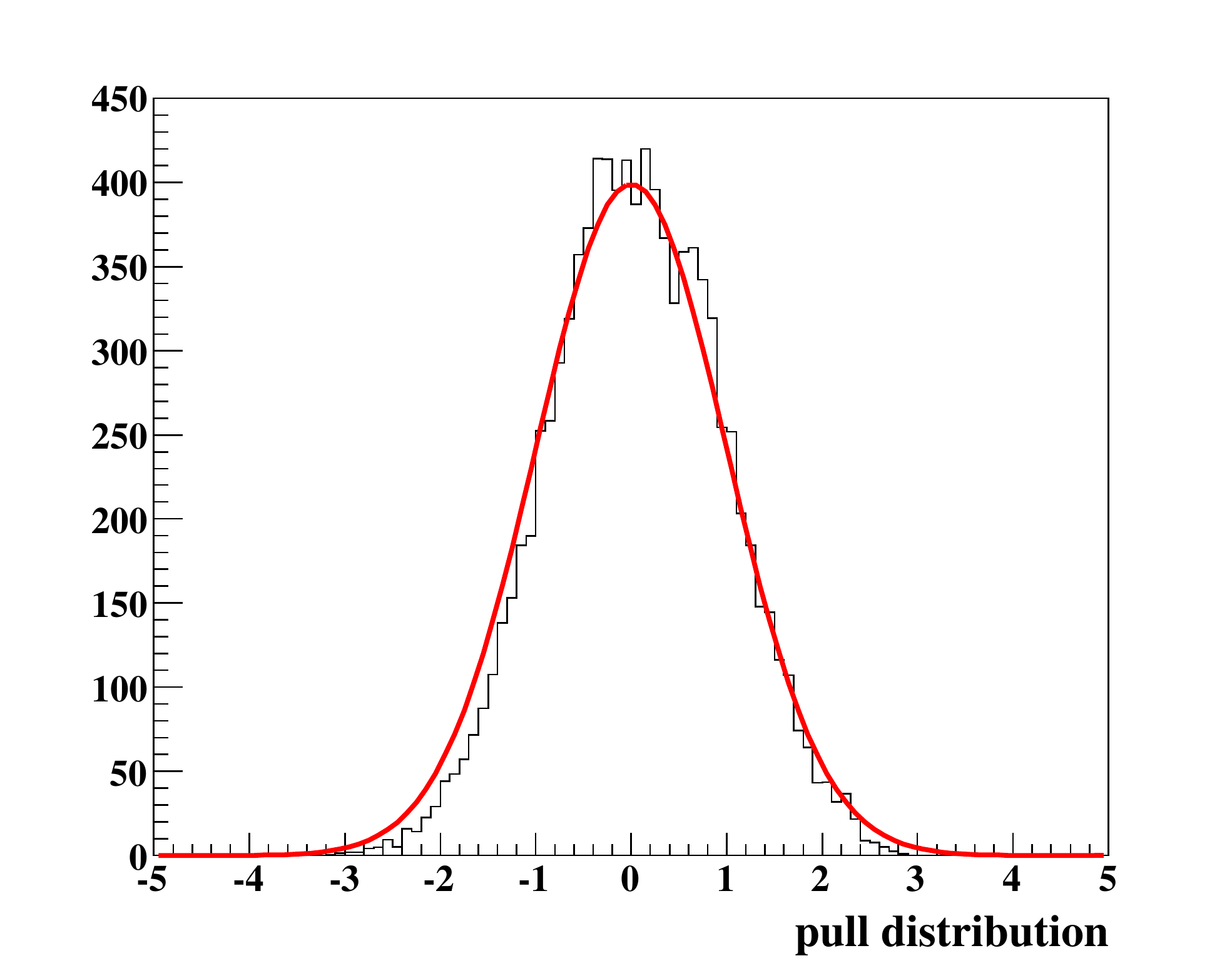}
  }
  \subfigure[]{
  \includegraphics[width=0.3\textwidth]{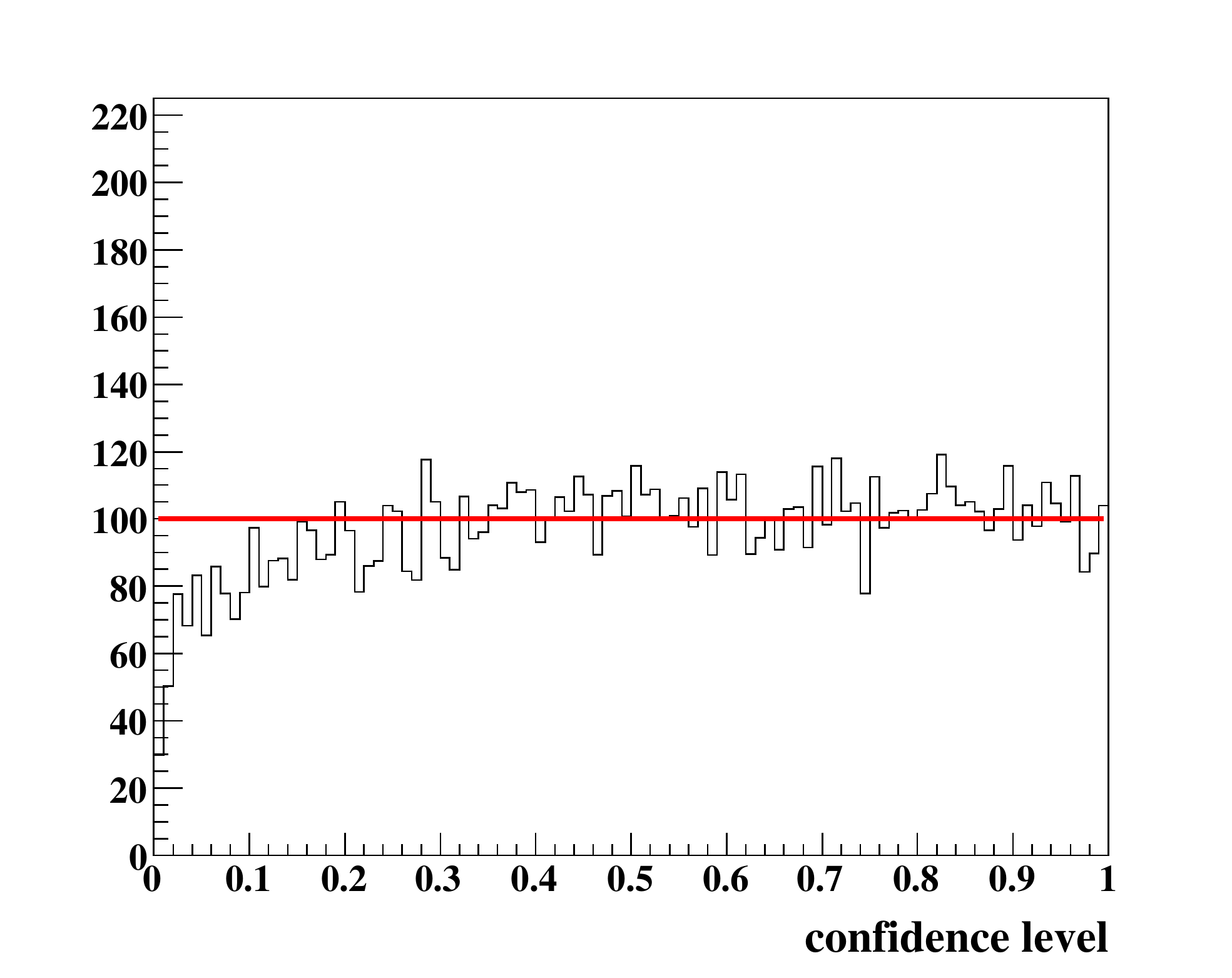}
  }
  \caption[]{\label{fig:chi2-cor-acc-bkgd}
    (Color Online) Events generated with background and with detector 
    acceptance given by (\ref{eq:detector-acc}).  
    (a) $\chi^2$ 
    (b) pull and 
    (c) confidence level distributions obtained using upper limits on the
    $Q$-factor errors (given in (\ref{eq:q-factor-upper-limit})). 
    The red lines represent the theoretical distributions, which contain no
    free parameters.
    See text for details.
  }
\end{figure*}

Calculating the correlation factor, $\rho^{jk}$, in 
(\ref{eq:n_m-error}) is very CPU intensive
due to the multiple loops which need to be performed over the data events. 
In~\cite{cite:sig-bkgd}, we noted that assuming 100\% correlation for all 
$Q$-factors provides a reasonable upper limit on the signal yield in any
kinematic region. 
Under this assumption, which eliminates much of the bookkeeping and CPU 
cycles needed to properly calculate  $\rho^{jk}$, the uncertainty on 
$n_{m_i}$ becomes
\begin{equation}
  \label{eq:q-factor-upper-limit}
  \sigma^2_{m_i} = n_{m_i} 
  + \left(\sum\limits_{j} \sigma_{Q_i}^j\right)^2,
\end{equation}
where the sum is again over the $n_c$ nearest neighbor events used to calculate
$Q_i$.

Figure~\ref{fig:chi2-cor-acc-bkgd} shows the $\chi^2$, pull and confidence 
level distributions obtained for our simulated data set under this assumption.
The deviations from the theoretical distributions are relatively small.
The total $\chi^2/n.d.f. = 8135.96/(9714.82-3) = 0.838$. 
This value is smaller than 1 as expected; however, it is still reasonably close
to the value obtained when the errors are handled rigorously. 
Thus, this is a reasonable approximation and could be used effectively in 
situations where calculating the values of $\rho^{jk}$ is not possible
(or, perhaps, desirable).

\subsection{\label{section:example:extend}Extending the Example}

To extend this example to allow for the case where the data is not binned in
production angle, we would simply need to include $\cos{\theta^{\omega}_{CM}}$
or $t$ in the vector of relevant coordinates, $\vec{\xi}$. 
To perform a full partial wave analysis on the data, we would also need to 
include any additional kinematic variables which factor into the partial wave 
amplitudes, {\em e.g.} the distance from the edge of the $\pi^+\pi^-\pi^0$ 
Dalitz plot (typically included in the $\omega$ decay amplitude). We would then
construct the likelihood from the partial waves and minimize 
$-\ln{\mathcal{L}}$ using the $Q$-factors obtained by applying our 
signal-background separation procedure, 
including the additional coordinates. 

The estimators from these event-based
fits would then serve as input into a higher dimensional version of the 
method presented in this example. The procedure for obtaining $z$-values, 
however, would remain almost unchanged. We would simply need to account for
the additional relevant kinematic variables in our metric.
If the fit provides a good 
description of the data, then the $\chi^2/n.d.f.$ should be approximately one
and the squares of the standardized residuals should follow a $\chi^2$
distribution.

\section{\label{section:conc}Conclusions}

In this paper, we have presented an {\em ad-hoc} procedure for obtaining
$\chi^2/n.d.f.$ values from unbinned maximum likelihood fits which does not
require binning the data. This makes it very applicable to multi-dimensional
problems. We have shown that these $\chi^2/n.d.f.$ values accurately reflect
how well a given hypothesis describes the data. 
\begin{center}
\mbox{}\\\vspace{0.2in}
{\normalsize \textbf{Acknowledgments}}
\end{center}
\vspace{0.2in}
This work was supported by grants from the United States Department of Energy
No. DE-FG02-87ER40315 and 
the National Science Foundation No. 0653316 through the 
``Physics at the Information Frontier'' program.

\end{document}